\documentclass[%
 aip,
 amsmath,amssymb,
preprint,%
superscriptaddress
]{revtex4-1}

\usepackage{graphicx}% Include figure files
\usepackage{dcolumn}% Align table columns on decimal point
\usepackage{bm}% bold math
\usepackage[utf8]{inputenc}
\usepackage[T1]{fontenc}
\usepackage{mathptmx}
\usepackage{etoolbox}
\usepackage{setspace}
\usepackage[colorlinks=true,linkcolor=blue,citecolor=blue,urlcolor=blue]{hyperref}
\makeatletter
\def\@email#1#2{%
 \endgroup
 \patchcmd{\titleblock@produce}
  {\frontmatter@RRAPformat}
  {\frontmatter@RRAPformat{\produce@RRAP{*#1\href{mailto:#2}{#2}}}\frontmatter@RRAPformat}
  {}{}
}%
\makeatother
\begin{document}

\preprint{}

\title[]{The role of disordered dynamics on the nature of transition in a turbulent reactive flow system}
% Force line breaks with \\
\author{Sivakumar Sudarsanan}
% \affiliation{%
%  Department of Aerospace Engineering, Indian Institute of Technology Madras, Chennai 600 036, India
% }%
% \affiliation{Centre of Excellence for studying Critical Transition in Complex Systems, Indian Institute of Technology Madras, Chennai  600 036, India}%Lines break automatically or can be forced 

\author{Induja Pavithran}%
% \email{}
% \affiliation{ 
% Authors' institution and/or address%\\This line break forced with \textbackslash\textbackslash
% }%

\author{R. I. Sujith}
\email{sujith@iitm.ac.in}
\affiliation{%
 Department of Aerospace Engineering, Indian Institute of Technology Madras, Chennai  600 036, India
}%
\affiliation{Centre of Excellence for Studying Critical Transition in Complex Systems, Indian Institute of Technology Madras, Chennai  600 036, India}  

\date{\today}% It is always \today, today,
             %  but any date may be explicitly specified

\begin{abstract}
The transition from a chaotic to a periodic oscillatory state can be smooth or abrupt in real-world turbulent systems. Although there have been several mathematical studies, the occurrence of abrupt transitions in real-world systems such as turbulent reactive flow systems is not well understood. A turbulent reactive flow system consists of the flame, the acoustic field, and the hydrodynamic field interacting nonlinearly. Generally, as the Reynolds number is increased, a laminar flow becomes turbulent, and the range of time scales associated with the flow broadens. Yet, as the Reynolds number is increased in a turbulent reactive flow system, a single dominant time scale emerges in the acoustic pressure oscillations, indicated by its loss of multifractality.  For such smooth and abrupt transitions from chaos to order, we study the evolution of correlated and uncorrelated dynamics between the acoustic pressure and the heat release rate oscillations in the spatiotemporal domain of the turbulent reactive system. The correlated dynamics that add or remove energy from the acoustic field are defined as conformists and contrarians, respectively. The uncorrelated dynamics, neither adds nor removes energy is defined as disorder. Conformist dynamics dominate the contrarian dynamics as order emerges from chaos. We discover that the spatial extent of the disordered dynamics plays a critical role in deciding the nature of the transition. During the smooth transition, we observe a significant presence of disordered dynamics in the spatial domain. In contrast, abrupt transitions are accompanied by the disappearance of disordered dynamics from the spatial domain.
\end{abstract}

\maketitle
% \begin{quotation}
% We study how smooth and abrupt transitions occur in a turbulent reactive flow system, which consists of the flame, the acoustic field, and the hydrodynamic flow field interacting in a nonlinear manner. We classify the dynamics between the acoustic pressure and the heat release rate oscillations that add or remove energy from the acoustic field as conformist and contrarian, while the dynamics that neither add nor remove energy as disorder. We discover that the spatial extent of the disordered dynamics plays a critical role in deciding the nature of the transition. During the smooth transition, we observe a significant presence of disordered dynamics in the spatial domain. In contrast, abrupt transitions are accompanied by the disappearance of disordered dynamics.
% \end{quotation}

\section{\label{sec:level1}Introduction}

The emergence of periodic oscillations or order from a state of chaos is intriguing; this transition can occur in a smooth or abrupt manner \cite{nair2014intermittency, Sudarsanan2024, singh2021intermittency, bhavi2023abrupt}. The amplitude of these periodic oscillations gradually increases during smooth, continuous phase transitions from chaos to order in far-from-equilibrium turbulent reactive flow systems \cite{nair2014intermittency, Sudarsanan2024}.  In contrast, an abrupt transition from chaos to order is associated with an explosive rise in the amplitude of the periodic oscillations with a hysteresis behavior \cite{singh2021intermittency, bhavi2023abrupt}. Abrupt transitions are more dangerous than smooth transitions. Examples of such abrupt transitions are tipping in climate systems \cite{van2016you}, epileptic seizures \cite{yaffe2015physiology}, and occurrence of violent high-amplitude periodic acoustic pressure oscillations in aircraft and rocket engines \cite{bhavi2023abrupt}.

\begin{figure*}%[H]
    \centering
    \includegraphics[width=\linewidth]{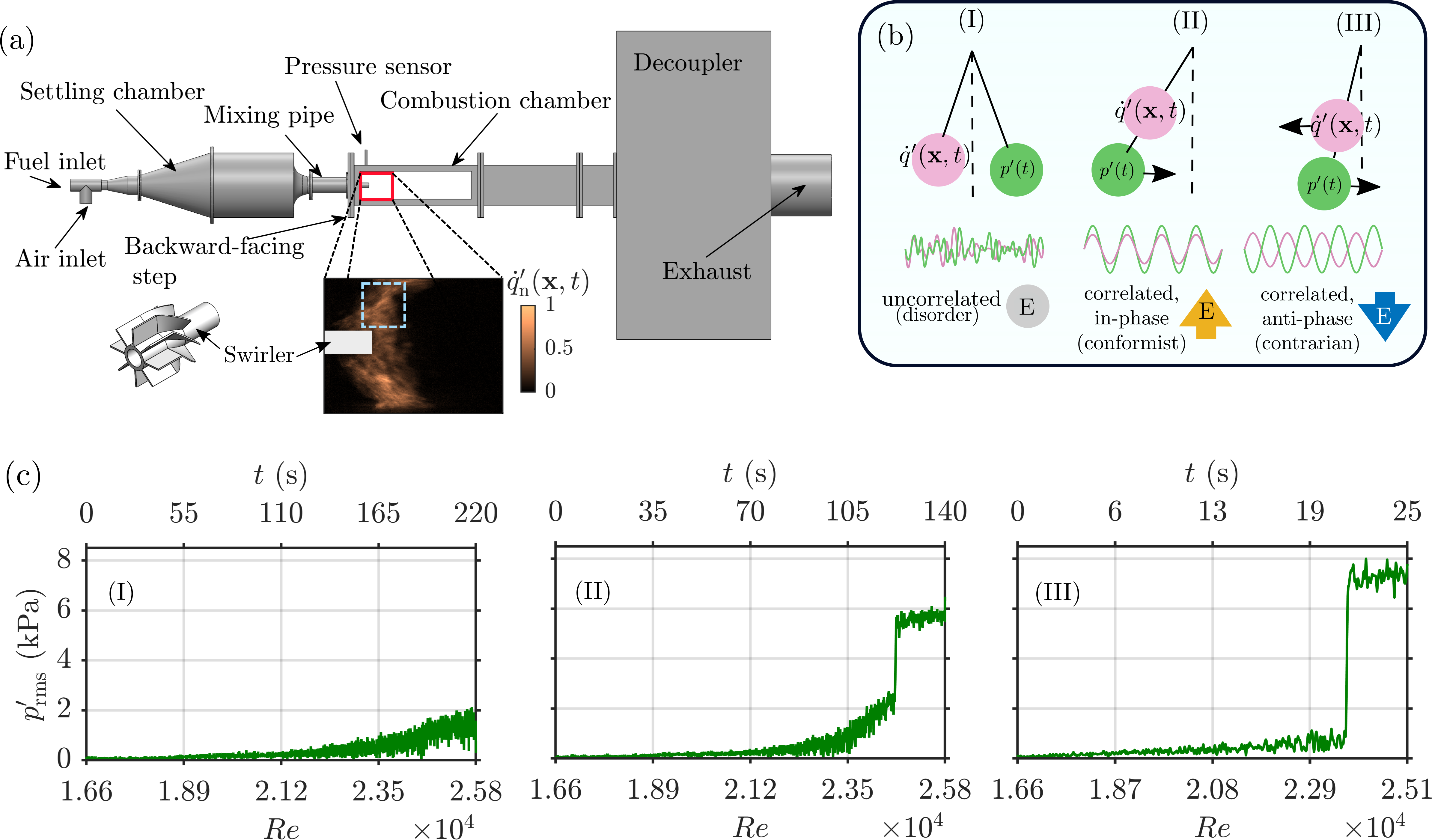}    
    \caption{A schematic of the experimental system consisting of air and fuel inlet ports, a settling chamber, a swirler used for flame stabilization, a combustion chamber, and a decoupler.  A piezoelectric pressure transducer attached to the top side of the combustion chamber is used to acquire the acoustic pressure measurements. The heat release rate dynamics (CH* images) were recorded using a high-speed camera. The zoomed-in view of the swirler and the combustion chamber are shown in (a).  The region inside the blue dashed rectangle in the inset of (a) is considered for the analysis. (b) We demonstrate the uncorrelated and correlated dynamics using the schematic of two coupled pendulums. The pink and the green pendulums represent the heat release rate ($\dot{q}^\prime(\mathbf{x},t)$) and the acoustic pressure ($p^\prime(t)$) oscillators. The same direction of the arrow marks for $\dot{q}^\prime(\mathbf{x},t)$ and $p^\prime(t)$ signifies in-phase oscillations, while the opposite direction indicates anti-phase oscillations. (b-I) The uncorrelated fluctuations neither add nor remove energy from the acoustic pressure field (disordered dynamics). (b-II) If $p^\prime (t)$ and $\dot{q}^\prime(\mathbf{x},t)$ oscillations are correlated and in-phase ($|\Delta \phi | < \pi/2$), energy will be added to the acoustic field (conformist dynamics). (b-III) If $p^\prime (t)$ and $\dot{q}^\prime (\mathbf{x},t)$ are correlated and anti-phase ($\pi/2 < | \Delta \phi | < \pi$), energy will be removed from the acoustic field (contrarian dynamics). (c) The evolution of the root-mean-square (rms) values of $p^\prime(t)$ calculated for a moving window of duration 0.5 s corresponding to (I) a smooth transition, (II) a smooth transition followed by an abrupt transition, and (III) an abrupt transition as the Reynolds number is continuously increased for three different experimental conditions.}
    \label{fig: transition}
    %\label{fig: experimentalSetup}
\end{figure*}

Abrupt transitions have been studied from the context of explosive synchronization and explosive percolation transition \cite{boccaletti2016explosive, d2019explosive, arola2022emergence}. Studies using mathematical models suggest that explosive synchronization transition arises in a network of phase oscillators under conditions of uniform frequency distribution of phase oscillators \cite{PazoThermodynamic}, positive correlation between the degree of an oscillator (degree is defined as the number of oscillators connected to a particular oscillator) and its frequency \cite{gomez2011explosive, leyva2013explosivephaseoscillators}, and higher-order interactions between the phase oscillators \cite{skardal2020higher,millan2020explosive}. Furthermore, studies show that the negatively coupled oscillators and positively coupled oscillators play a pivotal role in the occurrence of abrupt transition \cite{jalan2019inhibition, zhang2016suppressing, rathore2023synchronization}. Such positive and negative interactions are observed in neuronal \cite{soriano2008development, myung2015gaba}, social \cite{Yi2013}, and ecological \cite{giron2016synchronization} systems. In synchronization theory, the positively and negatively coupled oscillators are referred to as conformists and contrarians \cite{hong2011kuramoto}. Apart from several studies using mathematical models, only a few experimental studies on abrupt transitions have been reported \cite{kumar2015experimental,mahler2020experimental,taylor2009dynamical,leyva2012explosive,bhavi2023abrupt, cualuguaru2020first, singh2021intermittency}. In this paper, we present how smooth and abrupt phase transitions occur in an experimental turbulent reactive flow system (Fig.~\ref{fig: transition}a) that is far from equilibrium.

The turbulent reactive flow system comprises three interconnected subsystems:  the flame, the acoustic pressure, and the turbulent, hydrodynamic flow field. In our study, the control parameter, the Reynolds number is defined as the ratio of inertial to the viscous force. In fluid flow systems, as the Reynolds number is increased, a laminar fluid flow becomes turbulent and the range of time scales associated with the flow increases \cite{pope2001turbulent}. In contrast, as the Reynolds number is increased in a turbulent reactive flow system, we observe that the chaotic fluctuations characterized by a multifractal spectrum lose their multifractal nature, and a single dominant time scale (periodic oscillations) emerges in the acoustic pressure \cite{nair2014multifractality}. The emergence of periodic oscillations is due to the positive feedback loop arising from the nonlinear interaction between the subsystems \cite{sujith2020complex,juniper2018sensitivity}. The self-sustained periodic oscillatory state of large amplitude in turbulent reactive flow systems is known as thermoacoustic instability. Thermoacoustic instability is observed in aircraft engines, rocket engines and land-based gas turbine power plants \cite{biggs2009rocketdyne, lieuwen2005combustion, kriesels1995high}. The occurrence of such a large amplitude oscillatory state is undesirable, as it can lead to catastrophic system failures \cite{biggs2009rocketdyne, lieuwen2005combustion, kriesels1995high, green2006failure}.

The transition to the state of thermoacoustic instability in turbulent systems is manifested as the emergence of order from chaos \cite{sujith2020complex}. In turbulent systems, the transition from chaos to thermoacoustic instability was observed to be smooth via a state of intermittency characterized by epochs of periodic oscillations amidst aperiodic fluctuations \cite{nair2014intermittency}. The growth of amplitude of the acoustic pressure oscillations during this transition follows universal scaling relations with fractal and spectral measures of the acoustic pressure oscillations \cite{pavithran2020universality_inst, pavithran2020universality_spectral}. The presence of intermittency and the observed scaling relations helped in developing early warning measures for impending thermoacoustic instability \cite{nair2014multifractality}. During the state of chaos, small and disconnected patches of ordered regions (spatial locations exhibiting the correlated dynamics between the heat release rate and the acoustic pressure oscillations) are observed in the spatial domain \cite{Sudarsanan2024, krishnan2019emergence}. These small ordered regions merge and form a giant cluster spanning the entire spatial domain during the state of thermoacoustic instability \cite{Sudarsanan2024, krishnan2019emergence}. For a smooth transition, the distribution of the size of the patterns of ordered regions exhibits scaling laws that fall under the universality class of 2+1 directed percolation \cite{Sudarsanan2024}. Recently, abrupt transitions to a very high amplitude periodic oscillatory state in turbulent reactive flow systems have also been reported \cite{bhavi2023abrupt, singh2021intermittency, joseph2024explosive}. Such an abrupt switch to a high amplitude state is very dangerous in a practical scenario. {Significant studies on smooth transitions in turbulent reactive flow systems have been conducted; however, the occurrence of abrupt transitions in these systems has not been well studied.}

In this work, we study the dynamics between the heat release rate and the acoustic pressure oscillations in the spatiotemporal domain of the system during a (1) smooth transition, (2) a smooth followed by an abrupt transition, and  (3) an abrupt transition, as the Reynolds number is continuously increased for three different experimental conditions (Fig.~\ref{fig: transition}c) (refer to \textit{Materials and Methods} for more details on the experimental conditions). In this regard, first, we discuss the influence of the correlated and uncorrelated dynamics of the heat release rate and the acoustic pressure fluctuations on the acoustic energy of the system.

\section{Role of conformist, contrarian, and disordered dynamics}

The illustration in Fig.~\ref{fig: transition}b represents the influence of the uncorrelated and correlated dynamics between the heat release rate ($\dot{q}^\prime(\mathbf{x},t)$) and the acoustic pressure fluctuations ($p^\prime(t)$) on the acoustic energy of the reactive flow systems. The acoustic energy source is characterized by the correlation between the heat release rate and the acoustic pressure oscillations \cite{rayleigh1878explanation}, which can be expressed as
\begin{eqnarray}
\mathcal{R}=\int_{0}^{T} p^\prime(t)\dot{q}^\prime(\mathbf{x},t)dt.
\label{Eq: Ray}
\end{eqnarray}
Here, $T$ is the time period of the acoustic pressure oscillations \cite{poinsot2005theoretical}. During the uncorrelated dynamics (as shown in Fig.~$\ref{fig: transition}$b-I), where $\dot{q}^\prime(\mathbf{x},t)$ and $p^\prime(t)$ are desynchronized and exhibiting aperiodic oscillations, the correlation value ($\mathcal{R}$) is very minimal, and the energy is neither added nor removed from the acoustic field. Such uncorrelated dynamics signifies the underlying turbulent nature of the fluid flow. During the correlated dynamics as shown in Fig.~$\ref{fig: transition}$b-II and Fig.~$\ref{fig: transition}$b-III, energy is either added to or removed from the acoustic field depending upon the phase difference ($\Delta \phi$) between $\dot{q}^\prime(\mathbf{x},t)$ and $p^\prime(t)$. If $\dot{q}^\prime(\mathbf{x},t)$ and   $p^\prime(t)$ are in-phase ($| \Delta \phi| < \pi/2$), energy is added locally to the acoustic field (Fig.~\ref{fig: transition}b-II) \cite{rayleigh1878explanation, samaniego1993low}. On the other hand, if $\dot{q}^\prime(\mathbf{x},t)$ and $p^\prime(t)$ are anti-phase ($ \pi/2 < |\Delta \phi| < \pi$),  energy is removed locally from the acoustic field  (Fig.~\ref{fig: transition}b-III) \cite{rayleigh1878explanation, samaniego1993low}. The state of thermoacoustic instability arises when the 
aggregate energy added to the acoustic field from the entire spatial domain is greater than the net energy losses from the acoustic field \cite{poinsot2005theoretical}. 

In synchronization theory, the conformist oscillators promote global synchrony, whereas the contrarian oscillators hinder the emergence of synchrony due to their opposing behavior \cite{hong2011kuramoto}. 
The correlated in-phase dynamics adds energy to the acoustic field, whereas the correlated anti-phase dynamics removes energy from the acoustic field. Therefore, the influence of the correlated in-phase and the correlated anti-phase dynamics on the acoustic field is similar to the influence of conformist and contrarian oscillators on global synchrony. Using the cross-correlation measures, we classify the dynamics of $\dot{q}^\prime(\mathbf{x},t)$ and $p^\prime(t)$ as uncorrelated (disordered), correlated in-phase (conformist), and correlated anti-phase (contrarian) dynamics. A state variable ($S(\textbf{x},t)$) is introduced to represent the dynamics at a spatial location, represented as $S(\textbf{x},t) = [S_{\mathrm{conf}}, S_{\mathrm{cont}}, S_{\mathrm{dis}} ]$ (refer equations \ref{Eq-StateVariable 1}-\ref{Eq-StateVariable 3} in \textit{Materials and Methods}). Here, the subscript `conf', `cont', and `dis' represent the conformist, contrarian, and disordered dynamics respectively.

Next, we define order parameters for each of these dynamics, for example, the order parameter for the conformist dynamics represents its fraction in the spatial domain, denoted as, 
\begin{align}
     \rho_{\mathrm{conf}}(t)=\frac{1}{L^2}\sum_{\mathbf{x}} S_{\mathrm{conf}}(\mathbf{x},t),  \label{Eq-OP conf} 
\end{align}
where, $L^2$ is the size of the spatial domain. In a similar manner, order parameters are defined for the contrarian ($\rho_{\mathrm{cont}}$) and the disordered dynamics as well ($\rho_{\mathrm{dis}}$).  
\begin{align}
     \rho_{\mathrm{cont}}(t)=\frac{1}{L^2}\sum_{\mathbf{x}} S_{\mathrm{cont}}(\mathbf{x},t),   \label{Eq-OP cont} \\
     \rho_{\mathrm{dis}}(t)=\frac{1}{L^2}\sum_{\mathbf{x}} S_{\mathrm{dis}}(\mathbf{x},t).   \label{Eq-OP dis}
\end{align}
We study the variation of the order parameters for each of these dynamics for the smooth and abrupt transitions in a turbulent reactive flow system.

\section{Results}
\subsection{Significant presence of disordered dynamics in the spatial domain during the smooth transition}

\begin{figure*}
    \centering
    \includegraphics[width=11.4cm,height=11.4cm]{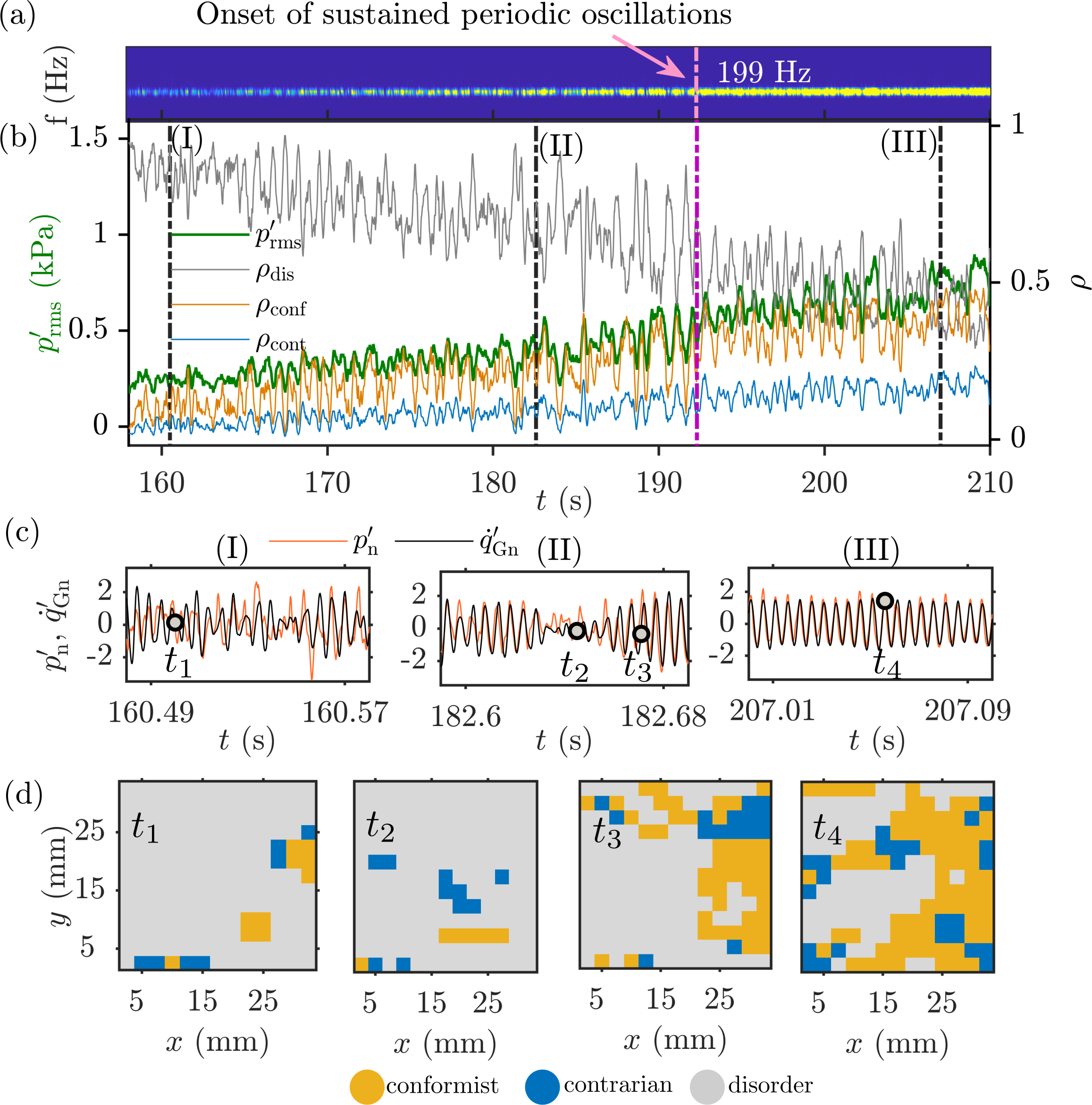}   
    \caption{({a}) Spectrogram of the acoustic pressure oscillations, (b) the variation of the rms values of the acoustic pressure fluctuations (green), and the evolution of the order parameters for the conformist (orange), contrarian (blue), and disordered dynamics (gray) corresponding to the smooth transition. The rms values are calculated for a moving window of duration 0.5 s.  The transition to the state of sustained periodic oscillations is evident in the spectrogram of the acoustic pressure oscillations and the rms values of the acoustic pressure oscillations gradually increase during this smooth transition. (c) The normalized $\dot{q}_{\mathrm{G}}^\prime(t)$ and the normalized $p^\prime(t)$ for the state of (I) low-amplitude aperiodic fluctuations, (II) intermittency, and (III) periodic oscillations for the epochs represented with black dashed lines in (b). (d) The spatial distribution of conformist, contrarian, and disordered dynamics correspond to the epochs $t_1$ to $t_4$ in (c). During the smooth transition, we observe the dominance of conformist dynamics over contrarian dynamics. This smooth transition is accompanied by a significant presence of disordered dynamics in the spatial domain.}
    \label{fig: smooth transition}
\end{figure*}

The spectrogram and the amplitude of the acoustic pressure oscillations (green color) during the smooth transition are shown in Fig.~\ref{fig: smooth transition}a, b respectively. As the control parameter, the Reynolds number ($Re$) is increased, a gradual rise in the amplitude of the acoustic pressure fluctuations is observed (Fig.~\ref{fig: smooth transition}b). The normalized acoustic pressure oscillations and the normalized heat release rate oscillations for different dynamical states during the smooth transition are shown in Fig.~\ref{fig: smooth transition}c.   The global heat release rate fluctuations ($\dot{q}_{\mathrm{G}}^\prime(t)$) are determined by summing the distribution of the local heat release rate oscillations which are obtained from the chemiluminescence imaging, i.e., $\dot{q}_{\mathrm{G}}^\prime(t)= \sum_{\mathbf{x}} \dot{q}^\prime(\mathbf{x},t)$. Throughout this paper, the normalized variables are denoted with a subscript `$\mathrm{n}$', and the variables are normalized using their respective standard deviation values. 

Initially ($Re=1.66 \times 10^4$), we observe low-amplitude aperiodic acoustic pressure oscillations which have been identified as high-dimensional chaos (Fig.~\ref{fig: smooth transition}b-I and Fig.~\ref{fig: smooth transition}c-I) \cite{tony2015detecting}. During the state of chaos, $\dot{q}_{\mathrm{Gn}}^\prime(t)$ and $p_{\mathrm{n}}^\prime(t)$ exhibit desynchronized dynamics  (Fig.~\ref{fig: smooth transition}c-I) \cite{pawar2017thermoacoustic}.
By increasing $Re$, we observe a smooth transition to a sustained, low-amplitude periodic oscillatory state, known as limit cycle oscillations (Fig.~\ref{fig: smooth transition}b-III and Fig.~\ref{fig: smooth transition}c-III). $\dot{q}_{\mathrm{Gn}}^\prime(t)$ and $p_{\mathrm{n}}^\prime(t)$ exhibit synchronized dynamics during the sustained periodic oscillatory state  (Fig.~\ref{fig: smooth transition}c-III) \cite{pawar2017thermoacoustic}. A state of intermittency where periodic oscillations appear amidst aperiodic oscillations is observed before the transition to the sustained periodic oscillatory state (Fig.~\ref{fig: smooth transition}c-II) \cite{nair2014intermittency}. The transition to the state of sustained periodic oscillations in the acoustic pressure is indicated by the appearance of a narrow, bright, continuous band in its spectrogram (Fig.~\ref{fig: smooth transition}a). The frequency of these periodic oscillations is 199 Hz, corresponding to the fundamental longitudinal acoustic mode of the combustion chamber.

Figure \ref{fig: smooth transition}d represents the spatial distribution of conformist (yellow), contrarian (blue), and disordered dynamics (grey) for four different epochs (labeled as $t_1$ to $t_4$ in  Fig.~\ref{fig: smooth transition}c) across the smooth transition. During the chaotic and the aperiodic epoch of the state of intermittency, the disordered dynamics is dominant in the spatial domain along with a negligible fraction of conformist and contrarian dynamics (Fig.~\ref{fig: smooth transition}d-$t_1$ and Fig.~\ref{fig: smooth transition}d-$t_2$). However, during the periodic epoch of the state of intermittency,  conformist and contrarian dynamics emerge in the spatial domain (Fig.~\ref{fig: smooth transition}d-$t_3$). After the transition to the state of sustained periodic oscillations (Fig.~\ref{fig: smooth transition}b-III), we observe that the fraction of regions exhibiting the conformist dynamics is greater than the fraction of regions exhibiting the contrarian dynamics (Fig.~\ref{fig: smooth transition}d-$t_4$) along with a significant spatial extent of the disordered dynamics that neither add nor remove energy from the acoustic field  (Fig.~\ref{fig: smooth transition}d-$t_4$). A greater fraction of conformist dynamics than contrarian dynamics ensures a continuous addition of energy to the acoustic field. 

The variation of order parameters for the conformist ($\rho_{\mathrm{conf}}$), contrarian ($\rho_{\mathrm{cont}}$), and disordered ($\rho_{\mathrm{dis}}$) dynamics across the smooth transition is shown in orange, blue, and grey colors respectively in Fig.~\ref{fig: smooth transition}b. During the smooth transition, we observe a gradual, monotonic increase of $\rho_{\mathrm{conf}}$ and $\rho_{\mathrm{cont}}$  from a very small value close to zero to values of 0.38 and 0.15 respectively at $t=210$ s (Fig.~\ref{fig: smooth transition}b). Meanwhile, the disordered dynamics  $\rho_{\mathrm{dis}}$, gradually decreases from 1 to 0.47 (Fig.~\ref{fig: smooth transition}b). Therefore, we observe that the conformist dynamics dominate over the contrarian dynamics, along with a significant spatial extent of disordered dynamics during the smooth transition from chaos to order (Fig.~\ref{fig: smooth transition}b).

\subsection{Disappearance of disordered dynamics from the spatial domain accompanying abrupt transitions}

Further, we consider two different scenarios: (1) a smooth transition followed by an abrupt transition and (2) an abrupt transition depicted in Fig.~\ref{fig: transition}c-II and Fig.~\ref{fig: transition}c-III, respectively.

\begin{figure*}[h!]
    \centering
    \includegraphics[width=\linewidth]{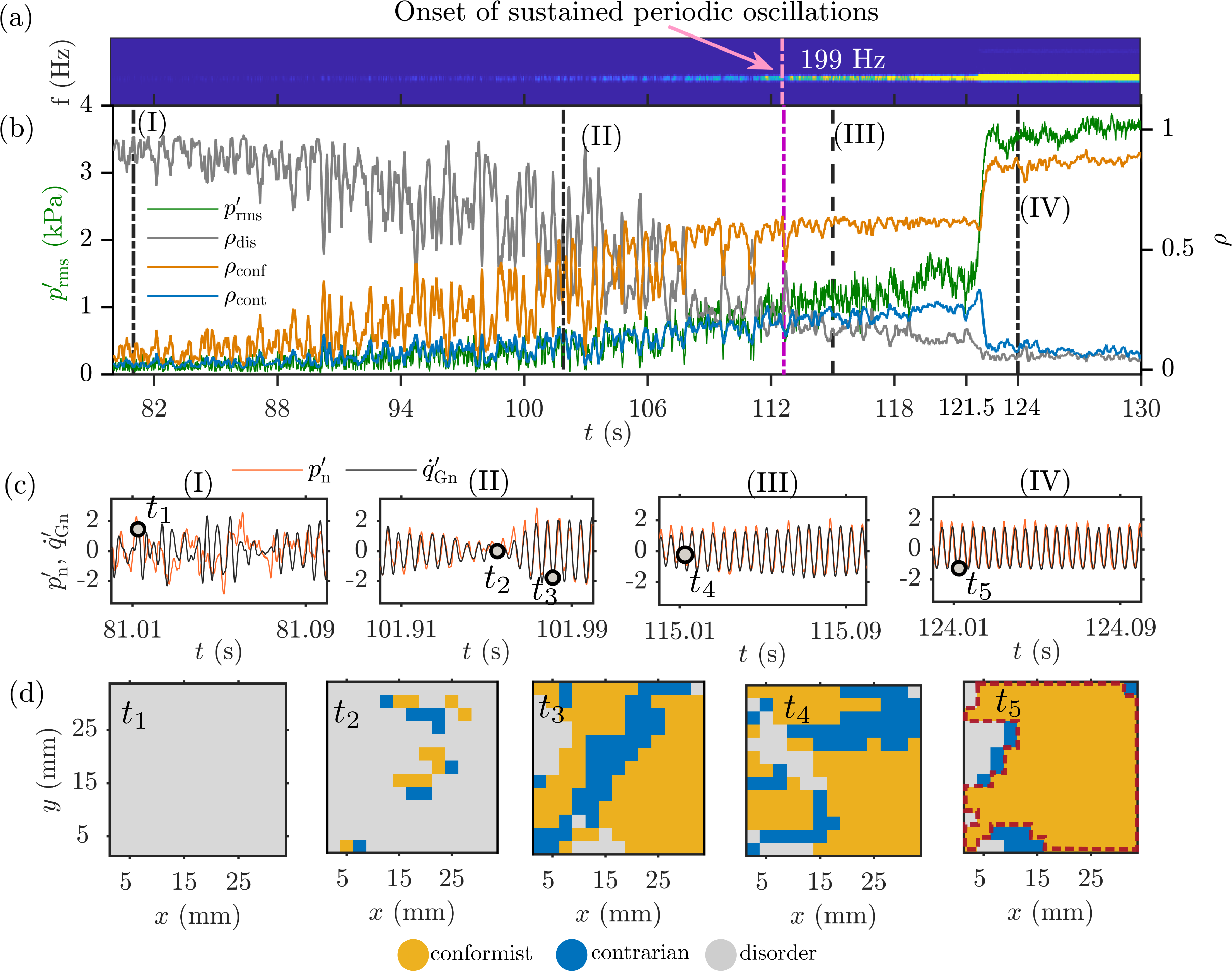}   
    \caption{({a}) The spectrogram and (b) the variation of the rms values of $p^\prime(t)$ for the smooth transition from a state of chaotic oscillations to a state of low-amplitude periodic oscillatory state followed by an abrupt transition to a high-amplitude periodic oscillatory state. (c) The normalized $\dot{q}_{\mathrm{G}}^\prime(t)$ and the normalized $p^\prime(t)$ for the state of (I) low-amplitude aperiodic oscillations, (II) intermittency, (III) low-amplitude periodic oscillatory state, (IV) high-amplitude periodic oscillatory state for the epochs marked with black dashed lines in (b). (d) The spatial distribution of conformist, contrarian, and disordered dynamics for the epochs labeled with $t_1$ to $t_5$ in (c). The evolution of the order parameters for the conformist, contrarian, and disordered dynamics during this transition is shown in (b). During the smooth transition, the conformist and contrarian dynamics emerge in the spatiotemporal domain. Further, an abrupt transition to a high-amplitude periodic oscillatory state is accompanied by an abrupt rise of $\rho_{\mathrm{conf}}$ and a transition to a state with minimal presence of disordered dynamics. A giant cluster exhibiting conformist dynamics, highlighted with a brown-dashed line in (d-$t_5$) emerges in the spatial domain accompanying the abrupt transition.}
    \label{fig: smooth followed by abrupt}
\end{figure*}

\begin{figure*}
    \centering
    \includegraphics[width=11.4cm,height=11.4cm]{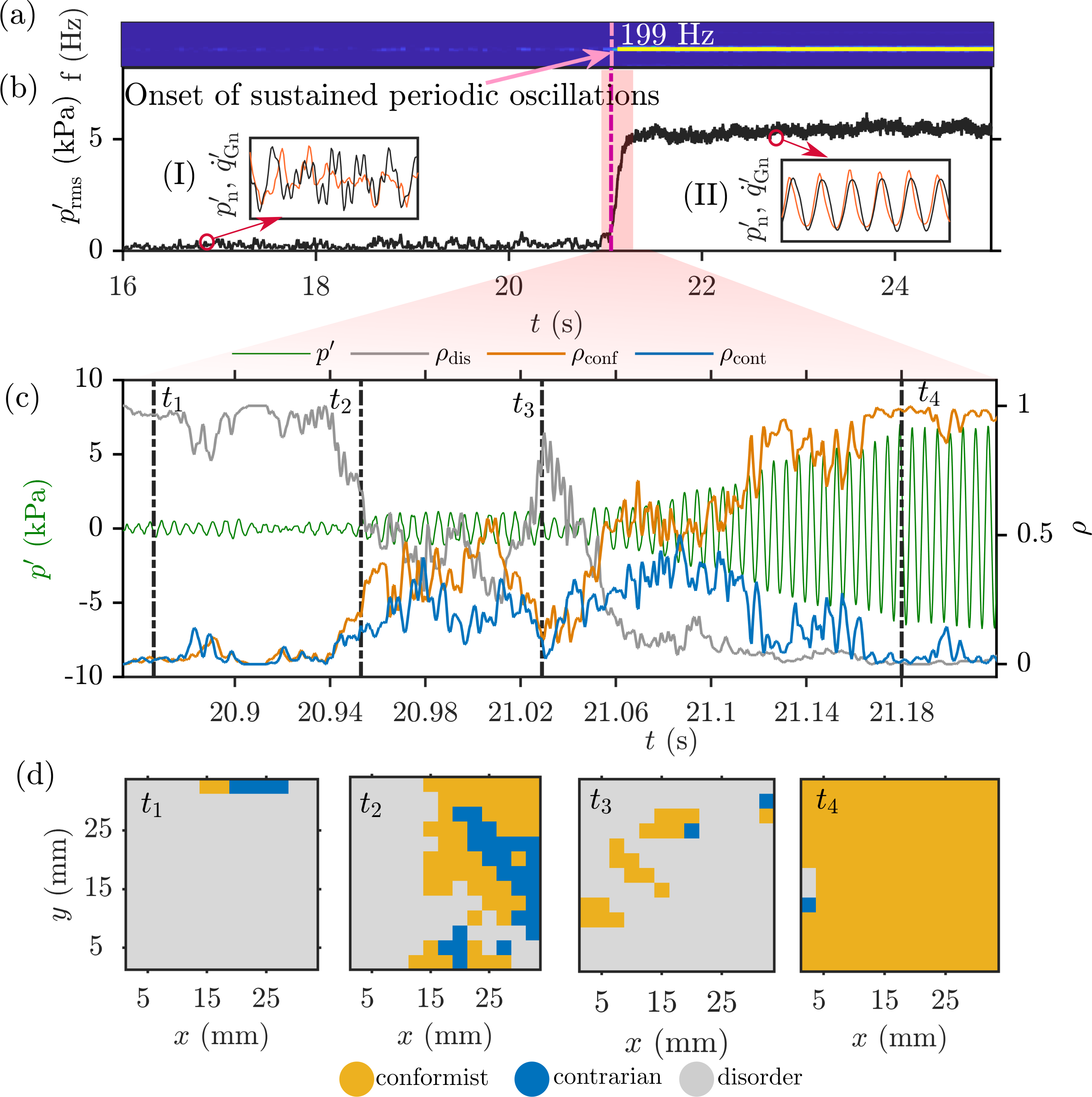}   
    \caption{({a}) The spectrogram and (b) the variation of the rms values of $p^\prime(t)$ for the abrupt transition from a state of chaotic fluctuations to a state of high-amplitude periodic oscillations. A small window of duration 0.5 s is considered for finding the rms of $p^\prime(t)$. The abrupt appearance of sustained periodic oscillations is evident in the spectrogram.  $p^\prime_{\mathrm{n}}(t)$ and $\dot{q}_{\mathrm{Gn}}^\prime(t)$ corresponding to the state of chaotic and high-amplitude periodic oscillations are shown in (b-I), and (b-II) respectively. (c) The zoomed-in view of  $p^\prime(t)$ and the evolution of the order parameters within the red box in (b). (d) The spatial distribution of conformist, contrarian, and disordered dynamics for the epochs $t_1$ to $t_4$ in (c). We find that the conformist dynamics dominate the contrarian dynamics after the abrupt transition. Accompanying this abrupt transition, the system transitions to a state with minimal disordered dynamics in the spatial domain.}
    \label{fig: abrupt transition}
\end{figure*}

For the first scenario, the spectrogram and the amplitude of the acoustic pressure oscillations are shown in Fig.~\ref{fig: smooth followed by abrupt}a, b. During this initial smooth transition, the amplitude of the acoustic pressure oscillations (green color line in Fig.~\ref{fig: smooth followed by abrupt}b) gradually increases while we observe a transition from a low-amplitude aperiodic oscillatory state (Fig.~\ref{fig: smooth followed by abrupt}b-I and Fig.~\ref{fig: smooth followed by abrupt}c-I) to a low-amplitude periodic oscillatory state (Fig.~\ref{fig: smooth followed by abrupt}b-III and Fig.~\ref{fig: smooth followed by abrupt}c-III) through a state of intermittency (Fig.~\ref{fig: smooth followed by abrupt}c-II). This transition from an aperiodic state to a state of sustained periodic oscillations is evident in the spectrogram of the acoustic pressure oscillations (Fig.~\ref{fig: smooth followed by abrupt}a). Further, this smooth transition is followed by an abrupt transition to a high-amplitude periodic oscillatory state  (Fig.~\ref{fig: smooth followed by abrupt}c-IV). 

During the low-amplitude aperiodic acoustic pressure oscillations, disordered dynamics dominate the spatial domain (Fig.~\ref{fig: smooth followed by abrupt}c and Fig.~\ref{fig: smooth followed by abrupt}d-$t_1$). We observe that the disordered dynamics are dominant during the aperiodic epoch of the state of intermittency as well (Fig.~\ref{fig: smooth followed by abrupt}d-$t_2$). However, during the periodic epoch of the state of intermittency, the conformist and contrarian dynamics emerge in the spatial domain (Fig.~\ref{fig: smooth followed by abrupt}d-$t_3$). During the low-amplitude periodic oscillatory state prior to the abrupt transition, we observe that medium-sized clusters of conformist dynamics are separated by clusters of contrarian and disordered dynamics (Fig.~\ref{fig: smooth followed by abrupt}d-$t_4$). After the abrupt transition, we observe that the contrarian and disordered dynamics separating the medium-sized clusters of conformist dynamics have disappeared, and a giant cluster of conformist dynamics that spans the entire spatial domain has emerged, which is highlighted with a brown-dashed line in Fig.~\ref{fig: smooth followed by abrupt}d-$t_5$.  

The evolution of the order parameters ($\rho_{\mathrm{conf}}$, $\rho_{\mathrm{cont}}$, and $\rho_{\mathrm{dis}}$) across the smooth transition followed by an abrupt transition is shown in Fig.~\ref{fig: smooth followed by abrupt}b. During the low-amplitude aperiodic oscillations (until $t=88$ s), the order parameters for both conformist and contrarian dynamics remain nearly zero ($\rho_{\mathrm{conf}} \approx 0$ and $\rho_{\mathrm{cont}} \approx 0$). During this state of aperiodic oscillations, the order parameter associated with disordered dynamics is close to one ($\rho_{\mathrm{dis}} \approx 1$). The values of $\rho_{\mathrm{conf}}$ and $\rho_{\mathrm{cont}}$  increases from 0 to  0.59 and 0.20 respectively between $t = 88$ s and $t=112$ s. Meanwhile, the disordered dynamics, $\rho_{\mathrm{dis}}$ decreases from 1 to 0.21. Between the interval (from $t=88$ s to $t=112$ s), we observe large fluctuations in $\rho_{\mathrm{conf}}$ and $\rho_{\mathrm{dis}}$ due to the appearance of intermittent periodic oscillations (Fig.~\ref{fig: smooth followed by abrupt}c-II). Furthermore, during the sustained low-amplitude periodic oscillatory state (between $t=112.6$ s and $t=121$ s), the values of $\rho_{\mathrm{conf}}$ and $\rho_{\mathrm{cont}}$  gradually increases to 0.63 and 0.27 respectively, meanwhile $\rho_{\mathrm{dis}}$  continues to decrease and reach a value of 0.1 at $t=121$ s.

After $t=121.5$ s, we observe an abrupt transition indicated by a steep rise in the amplitude of $p^\prime$ and an abrupt rise of $\rho_{\mathrm{conf}}$ (Fig.~\ref{fig: smooth followed by abrupt}b). The abrupt rise in the amplitude of $p^\prime$ coincides with an initial increase of $\rho_{\mathrm{cont}}$, but then $\rho_{\mathrm{cont}}$ decreases to a very small value during this abrupt transition (after $t=121.5$ s in Fig.~\ref{fig: smooth followed by abrupt}b). More importantly, we find that the system transitions to a state with minimal disordered dynamics accompanying this abrupt transition.  

For the second scenario, as illustrated in Fig.~\ref{fig: transition}c-III, we observe an abrupt transition from a state of low-amplitude aperiodic acoustic pressure oscillations (Fig.~\ref{fig: abrupt transition}b-I) to a state of high-amplitude periodic oscillatory state (Fig.~\ref{fig: abrupt transition}b-II). This abrupt transition is marked by the sudden appearance of a narrow, bright band in the spectrogram of the acoustic pressure oscillations indicated in Fig.~\ref{fig: abrupt transition}a. To understand the dynamics during the abrupt transition, consider the zoomed-in view of the epoch highlighted with red color in Fig.~\ref{fig: abrupt transition}b, which is shown in Fig.~\ref{fig: abrupt transition}c.

The evolution of order parameters and the spatial distribution of the conformist, contrarian, and disordered dynamics during the abrupt transition is shown in Fig.~\ref{fig: abrupt transition}c and Fig.~\ref{fig: abrupt transition}d, respectively. During the low-amplitude acoustic pressure fluctuations, the spatial regions exhibiting the disordered dynamics are very high as compared to the conformist and contrarian dynamics (Fig.~\ref{fig: abrupt transition}d-$t_1$). We observe the emergence of sustained periodic acoustic pressure oscillations after $t=20.94$ s onwards (Fig.~\ref{fig: abrupt transition}c). The amplitude of these periodic oscillations initially increases from $t=20.94$ s to $t=20.98$ s,  while $\rho_{\mathrm{conf}}$ and $\rho_{\mathrm{cont}}$ increase from 0 to 0.36 and 0.26 respectively. During this epoch (from $t=20.94$ s to $t=20.98$ s), $\rho_{\mathrm{dis}}$ decreases from 1 to 0.38. Further, the amplitude of the acoustic pressure oscillations does not grow monotonically, it decreases later (from $t=21.02$ s to $t=21.04$ s). Therefore we observe a non-monotonic variation of $\rho_{\mathrm{conf}}$, $\rho_{\mathrm{cont}}$ and $\rho_{\mathrm{dis}}$ between $t=21.02$ s and $t=21.04$ s (Fig.~\ref{fig: abrupt transition}c). $\rho_{\mathrm{dis}}$ increases first and then decreases whereas $\rho_{\mathrm{conf}}$ and $\rho_{\mathrm{cont}}$ decreases first and then increases. Further, $\rho_{\mathrm{dis}}$ continues to decrease and reach a small value of 0.15 around  $t=21.06$ s. Eventually, after $t=21.1$ s, we observe an abrupt transition indicated by a rapid rise of $p^\prime(t)$ and $\rho_{\mathrm{conf}}$ (Fig.~\ref{fig: abrupt transition}c). Accompanying this abrupt transition, we find that the system transitions to a state with minimal disordered dynamics (Fig.~\ref{fig: abrupt transition}c and Fig.~\ref{fig: abrupt transition}d-$t_4$).

\section{Discussions}

In a reactive flow system operating in turbulent conditions, we observe (1) a smooth transition, (2) a smooth transition followed by an abrupt transition, and (3) an abrupt transition from chaos to order for different experimental conditions (Fig.~\ref{fig: transition}c). We observe the dominance of conformist dynamics over contrarian dynamics during all three transitions we have discussed. The dominance of conformist dynamics maintains a flux of energy to the acoustic field during the state of sustained periodic oscillations. Notably, the evolution of disordered dynamics during the smooth and abrupt transitions differ. During the smooth transition (Fig.~\ref{fig: transition}c-I), we observe a significant spatial extent of disordered dynamics (Fig.~\ref{fig: smooth transition}b). Bhavi \textit{et al.} \cite{bhavi2023abrupt} reveals that turbulence intensity, in comparison to the amplitude of the limit cycle oscillations, smoothens the abrupt transition to the state of limit cycle oscillations. In the case of a smooth transition followed by an abrupt transition (Fig.~\ref{fig: transition}c-II), during the initial smooth transition, the disordered dynamics become minimal and continue to decrease, followed by an abrupt transition accompanied by the disappearance of disordered dynamics (Fig.~\ref{fig: smooth followed by abrupt}b). The abrupt transition illustrated in Fig.~\ref{fig: transition}c-III is accompanied by the disappearance of disordered dynamics (Fig.~\ref{fig: abrupt transition}c). In summary, we discover that the spatial extent of disordered dynamics plays a crucial role in deciding the nature of the transition. During the smooth transition, we observe a significant presence of disordered dynamics. In contrast, the abrupt transition is accompanied by the disappearance of disordered dynamics from the spatial domain.

In this paper, we have studied smooth and abrupt transitions from chaos to order by classifying various dynamics between the acoustic pressure and the heat release rate oscillations based on their influence on the acoustic energy of the system. Our methodology is inspired and based on the principles of synergetics and phase transition, wherein order parameters are defined for various plausible configurations to study the emergent behavior in complex systems \cite{haken1977synergetics}. Our study provides insights on the role of disordered dynamics on the occurrence of smooth and abrupt transitions in a real-world complex system.

\section{Materials and Methods}

\subsection{Experimental setup}
We performed experiments in a reactive flow system operating in turbulent conditions (Fig.~\ref{fig: transition}a). Our experimental setup consists of a settling chamber, mixing pipe, swirler, and combustion chamber of 800 mm in length. Air is first passed through a moisture separator and then flows into the settling chamber. We use liquified petroleum gas, composed of $40\%$ C$_3$H$_8$ and $60 \%$ C$_4$H$_{10}$ by volume, as fuel. The fuel is supplied into the mixing pipe through a central shaft with four radial injection holes, each with a diameter of 1.7 mm. These injection holes are located 100 mm upstream of the exit plane of the mixing pipe. A swirler is used to introduce radial momentum to the reactants flowing axially inside the mixing pipe. The swirler consists of eight fixed vanes of 1 mm thickness and 30 mm length. The fixed vane has a bent of $40^{\circ}$ about the axial direction of the combustion chamber. The swirler has a central body with a diameter of 16 mm and a length of 30 mm. The swirler is positioned in such a way that the downstream end of the swirler is aligned with the exit plane of the mixing pipe. The partially premixed air and fuel (reactants) inside the mixing pipe flow into the combustion chamber through a backward-facing step. The outlet of the combustion chamber is connected to a decoupler of size 1000 mm × 500 mm × 500 mm, which isolates the system from ambient fluctuations. 

The air and fuel flow rates are controlled by mass flow controllers (Alicat scientific MCR series). The mass flow controllers have an uncertainty of $\pm 0.8\%$ of the measured reading $+~0.2 \%$ of full-scale reading. The Reynolds number ($Re$) is defined as $\rho \bar{v} D/ \mu$, where $\mu$ and $\rho$ are the dynamic viscosity and density of the reactants calculated by considering the chemical composition of the reactants \cite{bird1961transport}, $D$ is the diameter of the swirler, and  $\bar{v}$ is the average velocity of the reactants at the backward-facing step of the combustion chamber. 

We conduct experiments by varying the flow rate of air from 7.76 g/s to 12.66 g/s for three different rates of 0.02, 0.03, and  0.2 g/s$^2$. The fuel flow rate is kept constant at 0.75 g/s during these experiments. This leads to the variation in the Reynolds number from $1.60 \times 10^4$ to $2.50 \times 10^4$ for three different rates: (1) $dRe/dt =340$ 1/s, (2) $dRe/dt =60$ 1/s and (3) $dRe/dt =38$ 1/s. Corresponding to these three different rates, we observe a smooth transition, a smooth transition followed by an abrupt transition, and an abrupt transition.

A piezoelectric pressure transducer (PCB103B02) is flush mounted to the wall of the combustion chamber at 120 mm downstream of the backward-facing step of the combustion chamber is used to measure the acoustic pressure oscillations inside the combustion chamber. The pressure transducer has an uncertainty of $\pm 0.15$ Pa and a sensitivity of 217.50 mV/kPa. The electrical signals from the piezoelectric pressure transducer were reordered at a sampling rate of 4000 Hz using a 16-bit analog-to-digital data acquisition system (NI-6143). The CH* chemiluminescence intensities represent the heat release rate fluctuations \cite{hardalupas2004local, guethe2012chemiluminescence}. A high-frame-rate camera (Phantom v12.1) with a Carl-Zeiss lens of 100 mm focal length, outfitted with a CH* filter, is used to record the chemiluminescence images. The outfitted CH* filter is characterized by a narrow bandwidth centered at 435 nm and has a full width at half maximum of 10 nm. The images were recorded at a sampling rate of 2000 Hz, simultaneously with the piezoelectric pressure transducer. The recorded images have a resolution of $512 \times 400$ pixels.

\subsection{Classifying the correlated dynamics}
In this section, we classify different correlated dynamics of the acoustic pressure fluctuations and the heat release rate fluctuations as shown in Fig.~\ref{fig: transition}b.  
The region where the flame is present is very small relative to the acoustic wavelength. Therefore, the acoustic pressure can be considered to be uniform within the reaction zone  \cite{Sudarsanan2024}.  We consider a linear correlation measure since the acoustic driving in reactive flow systems is proportional to the linear cross-correlation between the acoustic pressure fluctuations and the heat release rate fluctuations \citep{rayleigh1878explanation, schadow1992combustion, poinsot2005theoretical}. The cross-correlation between  $\dot{q}_{\mathrm{n}}^\prime(\mathbf{x},t)$ and $p_{\mathrm{n}}^\prime(t)$ for different time delay values are defined as

\begin{eqnarray}
\mathcal{R}(\mathbf{x},t,\tau_i)=\int_{t}^{t+W} p_{\mathrm{n}}^\prime(t_1)\dot{q}_{\mathrm{n}}^\prime(\mathbf{x},t_1+\tau_i)dt_1.
\label{eq: corr}
\end{eqnarray}
 
The cross-correlation values, $\mathcal{R}$, are calculated over a short duration ($W$) of $4T$, with $T$ representing the time period of the fundamental longitudinal mode of the acoustic oscillations. Our results are consistent for different short time window values  (\textit{Supporting  Information} Fig.~\ref{fig: windowSize}). $\mathcal{R}$ is calculated for distinct time delay values ($\tau_i$) from $-T/2$ to $+T/2$. 

\begin{figure}
\centering
\includegraphics[width=0.5\linewidth]{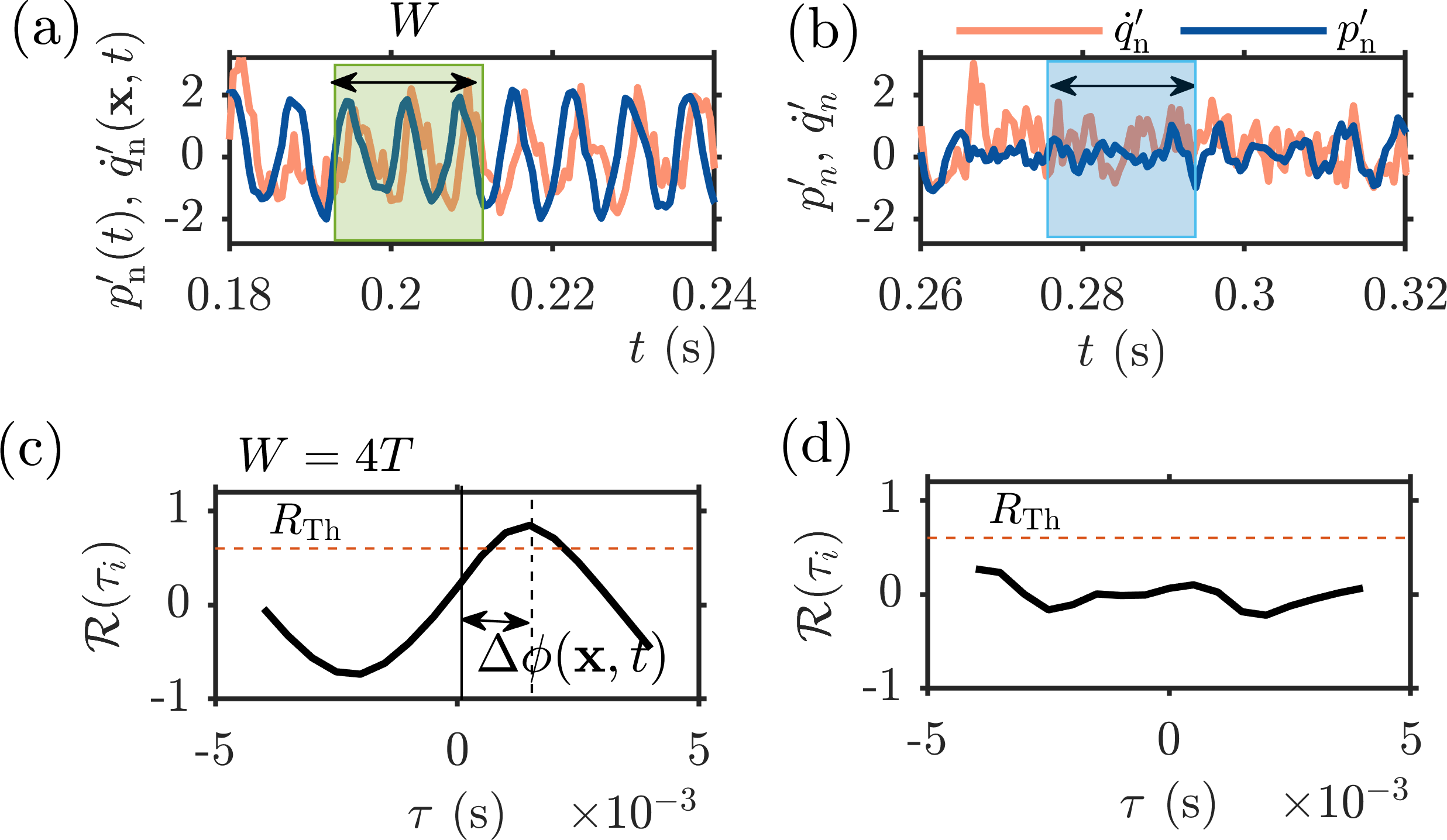}
\caption{
The normalized heat release rate oscillations (orange color) from a spatial location and the normalized acoustic pressure oscillations (blue color) for the state of (a) periodic and (b) aperodic oscillations.  (c, d) Cross-correlation values ($\mathcal{R}$) for a duration of $W=4T$  between $\dot{q}_{\mathrm{n}}^\prime$ and $p^\prime_{\mathrm{n}}$ for various lag values ($\tau_i$) for the epochs highlighted in (a, b). The orange dotted line is the threshold value for classifying the uncorrelated and correlated dynamics. For the uncorrelated dynamics, as shown in (d), the values of $\mathcal{R}$ remain very small for all $\tau$ values. However, for the correlated dynamics, $\mathcal{R}\approx 1$, when $\tau_i$ is equal to $\Delta \phi (\mathbf{x},t)$. Using this methodology, we distinguish the uncorrelated and the correlated dynamics. }
\label{Fig: sup R_vs_tau}
\end{figure}

\begin{figure}
    \centering
    \includegraphics[width=\linewidth]{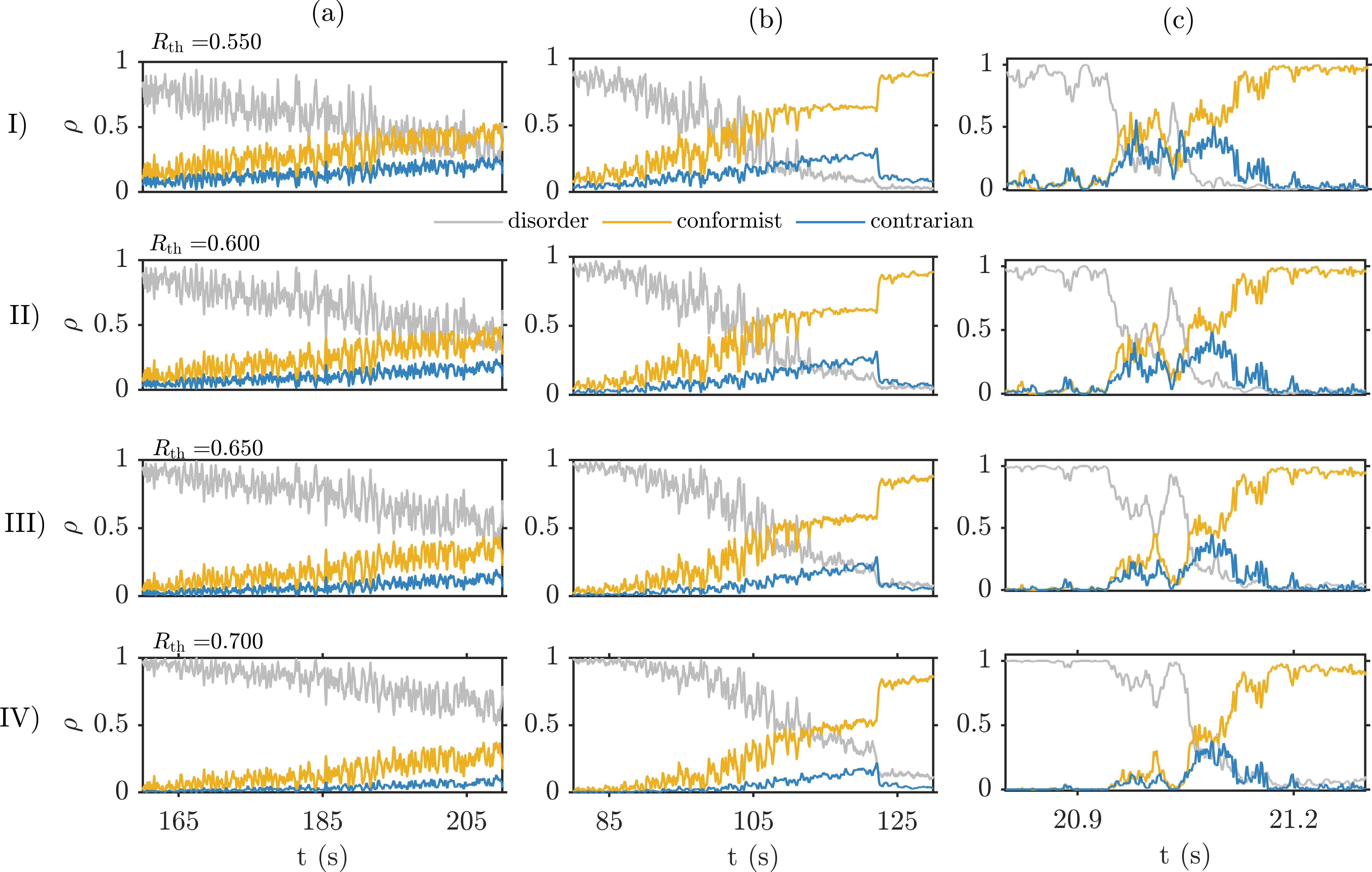}   
    \caption{
    {a-c}, The evolution of the order parameters for the conformist, contrarian, and disordered dynamics for different values of $R_{\mathrm{th}}=~0.55,~0.6,~0.65,$ and $0.7$. The evolution of order parameters remains qualitatively the same across (1) a smooth transition, (2) a smooth transition followed by an abrupt transition, and (3) an abrupt transition for a range of threshold values between 0.55 and 0.7.}
    \label{fig: threshold depedence}
\end{figure}

For the uncorrelated fluctuations in $\dot{q}^\prime(\mathbf{x},t)$ and $p^\prime(t)$, $\mathcal{R}(\mathbf{x},t)$ values remain small (Fig.~\ref{Fig: sup R_vs_tau}b, d), regardless of the time delay values.  In contrast, for the correlated fluctuations, a high correlation value ($\mathcal{R}(\mathbf{x},t)$) close to one is observed for a time lag value equivalent to the phase difference between $p^\prime(t)$ and $\dot{q}^\prime (\mathbf{x},t)$ (Fig.~\ref{Fig: sup R_vs_tau}a, c).
Finally, the maximum value among the correlation values for different time lag values is selected, expressed as 
\begin{equation}
R_{\mathrm{max}}(\mathbf{x},t)=\mathrm{max~for~all~}\tau_i \{\mathcal{R}(\mathbf{x},t,\tau_i)\}=\mathcal{R}(\mathbf{x},t,\Delta \phi(\mathbf{x},t)),  
\label{Eq-Rmax}
\end{equation}
where  $\Delta \phi(\mathbf{x},t)$ represents the phase difference between the variables, $\dot{q}_{\mathrm{n}}^\prime(\mathbf{x},t)$ and $p_{\mathrm{n}}^\prime(t)$. By selecting an appropriate threshold for $R_\mathrm{max}$, we can identify the uncorrelated and correlated dynamics in the spatiotemporal domain as illustrated in Fig.~\ref{Fig: sup R_vs_tau}. The threshold value for the $R_\mathrm{max}$ values is 0.6 for the present study. We have shown that our results qualitatively remain the same for a range of threshold values (refer to \textit{Supporting Information}, Fig.~\ref{fig: threshold depedence} for more details). The correlated dynamics with a time-shift value $| \Delta \phi(\mathbf{x},t)| \leq \pi/2$ are in-phase correlated dynamics (conformist) and correlated dynamics with $ \pi/2 < |\Delta \phi(\mathbf{x},t)| < \pi$ are anti-phase correlated dynamics (contrarian).

Having defined different corrected dynamics, a state variable, $S(\textbf{x},t) = [S_{\mathrm{conf}}, S_{\mathrm{cont}}, S_{\mathrm{dis}} ]$, is introduced to represent these correlated dynamics. Here, the subscript `conf', `cont', and `dis' represent the conformist, contrarian, and disordered dynamics, respectively. The state variable is defined as, 
\begin{align}
  S_{\mathrm{conf}}({\textbf{x},t}) &= 
    \begin{cases}
      1 & \text{if $R_{\mathrm{max}}(\textbf{x},t)\geq R_\text{Th}$ and } |\Delta \phi(\textbf{x},t)| \leq \pi/2, \\
      0 & \mathrm{else}.
    \end{cases}   \label{Eq-StateVariable 1} \\
  S_{\mathrm{cont}}({\textbf{x},t}) &=
    \begin{cases}
      1 & \text{if $R_{\mathrm{max}}(\textbf{x},t)\geq R_\text{Th}$ and } \pi/2 <|\Delta \phi(\textbf{x},t)| < \pi, \\
      0 & \mathrm{else}.
    \end{cases}   \label{Eq-StateVariable 2}   \\
  S_{\mathrm{dis}}({\textbf{x},t}) &=
    \begin{cases}
      1 & \text{if $R_{\mathrm{max}}(\textbf{x},t) < R_\text{Th}$}, \\
      0 & \mathrm{else}.
    \end{cases}  \label{Eq-StateVariable 3}      
\end{align}

\begin{acknowledgments}

We acknowledge Mrs. Shruti Tandon, Mr. Praveen Kumar, Mr. Ramesh S Bhavi, and Mr. Rohit R for their suggestions and feedback. We acknowledge Mr. P. R. Midhun, Mr. S. Anand, Mr. S. Thilagaraj, Mrs. G. Sudha and Dr. M. Ragunathan for their help during the experiments. RIS wishes to express his gratitude to the Department of Science and Technology and Ministry of Human Resource Development, Government of India for providing financial support for our research work under Grant No. JCB/2018/000034/SSC (JC Bose Fellowship).

\end{acknowledgments}

\bibliography{aipsamp}% Produces the bibliography via BibTeX.

%merlin.mbs aipnum4-1.bst 2010-07-25 4.21a (PWD, AO, DPC) hacked
%Control: key (0)
%Control: author (8) initials jnrlst
%Control: editor formatted (1) identically to author
%Control: production of article title (0) allowed
%Control: page (1) range
%Control: year (1) truncated
%Control: production of eprint (0) enabled
\begin{thebibliography}{49}%
\makeatletter
\providecommand \@ifxundefined [1]{%
 \@ifx{#1\undefined}
}%
\providecommand \@ifnum [1]{%
 \ifnum #1\expandafter \@firstoftwo
 \else \expandafter \@secondoftwo
 \fi
}%
\providecommand \@ifx [1]{%
 \ifx #1\expandafter \@firstoftwo
 \else \expandafter \@secondoftwo
 \fi
}%
\providecommand \natexlab [1]{#1}%
\providecommand \enquote  [1]{``#1''}%
\providecommand \bibnamefont  [1]{#1}%
\providecommand \bibfnamefont [1]{#1}%
\providecommand \citenamefont [1]{#1}%
\providecommand \href@noop [0]{\@secondoftwo}%
\providecommand \href [0]{\begingroup \@sanitize@url \@href}%
\providecommand \@href[1]{\@@startlink{#1}\@@href}%
\providecommand \@@href[1]{\endgroup#1\@@endlink}%
\providecommand \@sanitize@url [0]{\catcode `\\12\catcode `\$12\catcode `\&12\catcode `\#12\catcode `\^12\catcode `\_12\catcode `\%12\relax}%
\providecommand \@@startlink[1]{}%
\providecommand \@@endlink[0]{}%
\providecommand \url  [0]{\begingroup\@sanitize@url \@url }%
\providecommand \@url [1]{\endgroup\@href {#1}{\urlprefix }}%
\providecommand \urlprefix  [0]{URL }%
\providecommand \Eprint [0]{\href }%
\providecommand \doibase [0]{http://dx.doi.org/}%
\providecommand \selectlanguage [0]{\@gobble}%
\providecommand \bibinfo  [0]{\@secondoftwo}%
\providecommand \bibfield  [0]{\@secondoftwo}%
\providecommand \translation [1]{[#1]}%
\providecommand \BibitemOpen [0]{}%
\providecommand \bibitemStop [0]{}%
\providecommand \bibitemNoStop [0]{.\EOS\space}%
\providecommand \EOS [0]{\spacefactor3000\relax}%
\providecommand \BibitemShut  [1]{\csname bibitem#1\endcsname}%
\let\auto@bib@innerbib\@empty
%</preamble>
\bibitem [{\citenamefont {Nair}, \citenamefont {Thampi},\ and\ \citenamefont {Sujith}(2014)}]{nair2014intermittency}%
  \BibitemOpen
  \bibfield  {author} {\bibinfo {author} {\bibfnamefont {V.}~\bibnamefont {Nair}}, \bibinfo {author} {\bibfnamefont {G.}~\bibnamefont {Thampi}}, \ and\ \bibinfo {author} {\bibfnamefont {R.~I.}\ \bibnamefont {Sujith}},\ }\bibfield  {title} {\enquote {\bibinfo {title} {Intermittency route to thermoacoustic instability in turbulent combustors},}\ }\href@noop {} {\bibfield  {journal} {\bibinfo  {journal} {Journal of Fluid Mechanics}\ }\textbf {\bibinfo {volume} {756}},\ \bibinfo {pages} {470--487} (\bibinfo {year} {2014})}\BibitemShut {NoStop}%
\bibitem [{\citenamefont {Sudarsanan}\ \emph {et~al.}(2024)\citenamefont {Sudarsanan}, \citenamefont {Roy}, \citenamefont {Pavithran}, \citenamefont {Tandon},\ and\ \citenamefont {Sujith}}]{Sudarsanan2024}%
  \BibitemOpen
  \bibfield  {author} {\bibinfo {author} {\bibfnamefont {S.}~\bibnamefont {Sudarsanan}}, \bibinfo {author} {\bibfnamefont {A.}~\bibnamefont {Roy}}, \bibinfo {author} {\bibfnamefont {I.}~\bibnamefont {Pavithran}}, \bibinfo {author} {\bibfnamefont {S.}~\bibnamefont {Tandon}}, \ and\ \bibinfo {author} {\bibfnamefont {R.~I.}\ \bibnamefont {Sujith}},\ }\bibfield  {title} {\enquote {\bibinfo {title} {Emergence of order from chaos through a continuous phase transition in a turbulent reactive flow system},}\ }\href {\doibase 10.1103/PhysRevE.109.064214} {\bibfield  {journal} {\bibinfo  {journal} {Phys. Rev. E}\ }\textbf {\bibinfo {volume} {109}},\ \bibinfo {pages} {064214} (\bibinfo {year} {2024})}\BibitemShut {NoStop}%
\bibitem [{\citenamefont {Singh}\ \emph {et~al.}(2021)\citenamefont {Singh}, \citenamefont {Roy}, \citenamefont {Reeja}, \citenamefont {Nair}, \citenamefont {Chaudhuri},\ and\ \citenamefont {Sujith}}]{singh2021intermittency}%
  \BibitemOpen
  \bibfield  {author} {\bibinfo {author} {\bibfnamefont {S.}~\bibnamefont {Singh}}, \bibinfo {author} {\bibfnamefont {A.}~\bibnamefont {Roy}}, \bibinfo {author} {\bibfnamefont {K.~V.}\ \bibnamefont {Reeja}}, \bibinfo {author} {\bibfnamefont {A.}~\bibnamefont {Nair}}, \bibinfo {author} {\bibfnamefont {S.}~\bibnamefont {Chaudhuri}}, \ and\ \bibinfo {author} {\bibfnamefont {R.~I.}\ \bibnamefont {Sujith}},\ }\bibfield  {title} {\enquote {\bibinfo {title} {Intermittency, secondary bifurcation and mixed-mode oscillations in a swirl-stabilized annular combustor: Experiments and modeling},}\ }\href@noop {} {\bibfield  {journal} {\bibinfo  {journal} {Journal of Engineering for Gas Turbines and Power}\ }\textbf {\bibinfo {volume} {143}},\ \bibinfo {pages} {051028} (\bibinfo {year} {2021})}\BibitemShut {NoStop}%
\bibitem [{\citenamefont {Bhavi}\ \emph {et~al.}(2023)\citenamefont {Bhavi}, \citenamefont {Pavithran}, \citenamefont {Roy},\ and\ \citenamefont {Sujith}}]{bhavi2023abrupt}%
  \BibitemOpen
  \bibfield  {author} {\bibinfo {author} {\bibfnamefont {R.~S.}\ \bibnamefont {Bhavi}}, \bibinfo {author} {\bibfnamefont {I.}~\bibnamefont {Pavithran}}, \bibinfo {author} {\bibfnamefont {A.}~\bibnamefont {Roy}}, \ and\ \bibinfo {author} {\bibfnamefont {R.~I.}\ \bibnamefont {Sujith}},\ }\bibfield  {title} {\enquote {\bibinfo {title} {Abrupt transitions in turbulent thermoacoustic systems},}\ }\href@noop {} {\bibfield  {journal} {\bibinfo  {journal} {Journal of Sound and Vibration}\ }\textbf {\bibinfo {volume} {547}},\ \bibinfo {pages} {117478} (\bibinfo {year} {2023})}\BibitemShut {NoStop}%
\bibitem [{\citenamefont {Van~Nes}\ \emph {et~al.}(2016)\citenamefont {Van~Nes}, \citenamefont {Arani}, \citenamefont {Staal}, \citenamefont {van~der Bolt}, \citenamefont {Flores}, \citenamefont {Bathiany},\ and\ \citenamefont {Scheffer}}]{van2016you}%
  \BibitemOpen
  \bibfield  {author} {\bibinfo {author} {\bibfnamefont {E.~H.}\ \bibnamefont {Van~Nes}}, \bibinfo {author} {\bibfnamefont {B.~M.~S.}\ \bibnamefont {Arani}}, \bibinfo {author} {\bibfnamefont {A.}~\bibnamefont {Staal}}, \bibinfo {author} {\bibfnamefont {B.}~\bibnamefont {van~der Bolt}}, \bibinfo {author} {\bibfnamefont {B.~M.}\ \bibnamefont {Flores}}, \bibinfo {author} {\bibfnamefont {S.}~\bibnamefont {Bathiany}}, \ and\ \bibinfo {author} {\bibfnamefont {M.}~\bibnamefont {Scheffer}},\ }\bibfield  {title} {\enquote {\bibinfo {title} {What do you mean,‘tipping point’?}}\ }\href@noop {} {\bibfield  {journal} {\bibinfo  {journal} {Trends in Ecology \& Evolution}\ }\textbf {\bibinfo {volume} {31}},\ \bibinfo {pages} {902--904} (\bibinfo {year} {2016})}\BibitemShut {NoStop}%
\bibitem [{\citenamefont {Yaffe}\ \emph {et~al.}(2015)\citenamefont {Yaffe}, \citenamefont {Borger}, \citenamefont {Megevand}, \citenamefont {Groppe}, \citenamefont {Kramer}, \citenamefont {Chu}, \citenamefont {Santaniello}, \citenamefont {Meisel}, \citenamefont {Mehta},\ and\ \citenamefont {Sarma}}]{yaffe2015physiology}%
  \BibitemOpen
  \bibfield  {author} {\bibinfo {author} {\bibfnamefont {R.~B.}\ \bibnamefont {Yaffe}}, \bibinfo {author} {\bibfnamefont {P.}~\bibnamefont {Borger}}, \bibinfo {author} {\bibfnamefont {P.}~\bibnamefont {Megevand}}, \bibinfo {author} {\bibfnamefont {D.~M.}\ \bibnamefont {Groppe}}, \bibinfo {author} {\bibfnamefont {M.~A.}\ \bibnamefont {Kramer}}, \bibinfo {author} {\bibfnamefont {C.~J.}\ \bibnamefont {Chu}}, \bibinfo {author} {\bibfnamefont {S.}~\bibnamefont {Santaniello}}, \bibinfo {author} {\bibfnamefont {C.}~\bibnamefont {Meisel}}, \bibinfo {author} {\bibfnamefont {A.~D.}\ \bibnamefont {Mehta}}, \ and\ \bibinfo {author} {\bibfnamefont {S.~V.}\ \bibnamefont {Sarma}},\ }\bibfield  {title} {\enquote {\bibinfo {title} {Physiology of functional and effective networks in epilepsy},}\ }\href@noop {} {\bibfield  {journal} {\bibinfo  {journal} {Clinical Neurophysiology}\ }\textbf {\bibinfo {volume} {126}},\ \bibinfo {pages} {227--236} (\bibinfo {year} {2015})}\BibitemShut {NoStop}%
\bibitem [{\citenamefont {Boccaletti}\ \emph {et~al.}(2016)\citenamefont {Boccaletti}, \citenamefont {Almendral}, \citenamefont {Guan}, \citenamefont {Leyva}, \citenamefont {Liu}, \citenamefont {Sendi{\~n}a-Nadal}, \citenamefont {Wang},\ and\ \citenamefont {Zou}}]{boccaletti2016explosive}%
  \BibitemOpen
  \bibfield  {author} {\bibinfo {author} {\bibfnamefont {S.}~\bibnamefont {Boccaletti}}, \bibinfo {author} {\bibfnamefont {J.}~\bibnamefont {Almendral}}, \bibinfo {author} {\bibfnamefont {S.}~\bibnamefont {Guan}}, \bibinfo {author} {\bibfnamefont {I.}~\bibnamefont {Leyva}}, \bibinfo {author} {\bibfnamefont {Z.}~\bibnamefont {Liu}}, \bibinfo {author} {\bibfnamefont {I.}~\bibnamefont {Sendi{\~n}a-Nadal}}, \bibinfo {author} {\bibfnamefont {Z.}~\bibnamefont {Wang}}, \ and\ \bibinfo {author} {\bibfnamefont {Y.}~\bibnamefont {Zou}},\ }\bibfield  {title} {\enquote {\bibinfo {title} {Explosive transitions in complex networks’ structure and dynamics: Percolation and synchronization},}\ }\href@noop {} {\bibfield  {journal} {\bibinfo  {journal} {Physics Reports}\ }\textbf {\bibinfo {volume} {660}},\ \bibinfo {pages} {1--94} (\bibinfo {year} {2016})}\BibitemShut {NoStop}%
\bibitem [{\citenamefont {D'Souza}\ \emph {et~al.}(2019)\citenamefont {D'Souza}, \citenamefont {G{\'o}mez-Gardenes}, \citenamefont {Nagler},\ and\ \citenamefont {Arenas}}]{d2019explosive}%
  \BibitemOpen
  \bibfield  {author} {\bibinfo {author} {\bibfnamefont {R.~M.}\ \bibnamefont {D'Souza}}, \bibinfo {author} {\bibfnamefont {J.}~\bibnamefont {G{\'o}mez-Gardenes}}, \bibinfo {author} {\bibfnamefont {J.}~\bibnamefont {Nagler}}, \ and\ \bibinfo {author} {\bibfnamefont {A.}~\bibnamefont {Arenas}},\ }\bibfield  {title} {\enquote {\bibinfo {title} {Explosive phenomena in complex networks},}\ }\href@noop {} {\bibfield  {journal} {\bibinfo  {journal} {Advances in Physics}\ }\textbf {\bibinfo {volume} {68}},\ \bibinfo {pages} {123--223} (\bibinfo {year} {2019})}\BibitemShut {NoStop}%
\bibitem [{\citenamefont {Arola-Fern{\'a}ndez}\ \emph {et~al.}(2022)\citenamefont {Arola-Fern{\'a}ndez}, \citenamefont {Faci-L{\'a}zaro}, \citenamefont {Skardal}, \citenamefont {Boghiu}, \citenamefont {G{\'o}mez-Garde{\~n}es},\ and\ \citenamefont {Arenas}}]{arola2022emergence}%
  \BibitemOpen
  \bibfield  {author} {\bibinfo {author} {\bibfnamefont {L.}~\bibnamefont {Arola-Fern{\'a}ndez}}, \bibinfo {author} {\bibfnamefont {S.}~\bibnamefont {Faci-L{\'a}zaro}}, \bibinfo {author} {\bibfnamefont {P.~S.}\ \bibnamefont {Skardal}}, \bibinfo {author} {\bibfnamefont {E.~C.}\ \bibnamefont {Boghiu}}, \bibinfo {author} {\bibfnamefont {J.}~\bibnamefont {G{\'o}mez-Garde{\~n}es}}, \ and\ \bibinfo {author} {\bibfnamefont {A.}~\bibnamefont {Arenas}},\ }\bibfield  {title} {\enquote {\bibinfo {title} {Emergence of explosive synchronization bombs in networks of oscillators},}\ }\href@noop {} {\bibfield  {journal} {\bibinfo  {journal} {Communications Physics}\ }\textbf {\bibinfo {volume} {5}},\ \bibinfo {pages} {264} (\bibinfo {year} {2022})}\BibitemShut {NoStop}%
\bibitem [{\citenamefont {Paz\'o}(2005)}]{PazoThermodynamic}%
  \BibitemOpen
  \bibfield  {author} {\bibinfo {author} {\bibfnamefont {D.}~\bibnamefont {Paz\'o}},\ }\bibfield  {title} {\enquote {\bibinfo {title} {Thermodynamic limit of the first-order phase transition in the kuramoto model},}\ }\href {\doibase 10.1103/PhysRevE.72.046211} {\bibfield  {journal} {\bibinfo  {journal} {Phys. Rev. E}\ }\textbf {\bibinfo {volume} {72}},\ \bibinfo {pages} {046211} (\bibinfo {year} {2005})}\BibitemShut {NoStop}%
\bibitem [{\citenamefont {G{\'o}mez-Gardenes}\ \emph {et~al.}(2011)\citenamefont {G{\'o}mez-Gardenes}, \citenamefont {G{\'o}mez}, \citenamefont {Arenas},\ and\ \citenamefont {Moreno}}]{gomez2011explosive}%
  \BibitemOpen
  \bibfield  {author} {\bibinfo {author} {\bibfnamefont {J.}~\bibnamefont {G{\'o}mez-Gardenes}}, \bibinfo {author} {\bibfnamefont {S.}~\bibnamefont {G{\'o}mez}}, \bibinfo {author} {\bibfnamefont {A.}~\bibnamefont {Arenas}}, \ and\ \bibinfo {author} {\bibfnamefont {Y.}~\bibnamefont {Moreno}},\ }\bibfield  {title} {\enquote {\bibinfo {title} {Explosive synchronization transitions in scale-free networks},}\ }\href@noop {} {\bibfield  {journal} {\bibinfo  {journal} {Physical Review Letters}\ }\textbf {\bibinfo {volume} {106}},\ \bibinfo {pages} {128701} (\bibinfo {year} {2011})}\BibitemShut {NoStop}%
\bibitem [{\citenamefont {Leyva}\ \emph {et~al.}(2013)\citenamefont {Leyva}, \citenamefont {Navas}, \citenamefont {Sendi{\~n}a-Nadal}, \citenamefont {Almendral}, \citenamefont {Buld{\'u}}, \citenamefont {Zanin}, \citenamefont {Papo},\ and\ \citenamefont {Boccaletti}}]{leyva2013explosivephaseoscillators}%
  \BibitemOpen
  \bibfield  {author} {\bibinfo {author} {\bibfnamefont {I.}~\bibnamefont {Leyva}}, \bibinfo {author} {\bibfnamefont {A.}~\bibnamefont {Navas}}, \bibinfo {author} {\bibfnamefont {I.}~\bibnamefont {Sendi{\~n}a-Nadal}}, \bibinfo {author} {\bibfnamefont {J.}~\bibnamefont {Almendral}}, \bibinfo {author} {\bibfnamefont {J.}~\bibnamefont {Buld{\'u}}}, \bibinfo {author} {\bibfnamefont {M.}~\bibnamefont {Zanin}}, \bibinfo {author} {\bibfnamefont {D.}~\bibnamefont {Papo}}, \ and\ \bibinfo {author} {\bibfnamefont {S.}~\bibnamefont {Boccaletti}},\ }\bibfield  {title} {\enquote {\bibinfo {title} {Explosive transitions to synchronization in networks of phase oscillators},}\ }\href@noop {} {\bibfield  {journal} {\bibinfo  {journal} {Scientific Reports}\ }\textbf {\bibinfo {volume} {3}},\ \bibinfo {pages} {1281} (\bibinfo {year} {2013})}\BibitemShut {NoStop}%
\bibitem [{\citenamefont {Skardal}\ and\ \citenamefont {Arenas}(2020)}]{skardal2020higher}%
  \BibitemOpen
  \bibfield  {author} {\bibinfo {author} {\bibfnamefont {P.~S.}\ \bibnamefont {Skardal}}\ and\ \bibinfo {author} {\bibfnamefont {A.}~\bibnamefont {Arenas}},\ }\bibfield  {title} {\enquote {\bibinfo {title} {Higher order interactions in complex networks of phase oscillators promote abrupt synchronization switching},}\ }\href@noop {} {\bibfield  {journal} {\bibinfo  {journal} {Communications Physics}\ }\textbf {\bibinfo {volume} {3}},\ \bibinfo {pages} {218} (\bibinfo {year} {2020})}\BibitemShut {NoStop}%
\bibitem [{\citenamefont {Mill{\'a}n}, \citenamefont {Torres},\ and\ \citenamefont {Bianconi}(2020)}]{millan2020explosive}%
  \BibitemOpen
  \bibfield  {author} {\bibinfo {author} {\bibfnamefont {A.~P.}\ \bibnamefont {Mill{\'a}n}}, \bibinfo {author} {\bibfnamefont {J.~J.}\ \bibnamefont {Torres}}, \ and\ \bibinfo {author} {\bibfnamefont {G.}~\bibnamefont {Bianconi}},\ }\bibfield  {title} {\enquote {\bibinfo {title} {Explosive higher-order \uppercase{K}uramoto dynamics on simplicial complexes},}\ }\href@noop {} {\bibfield  {journal} {\bibinfo  {journal} {Physical Review Letters}\ }\textbf {\bibinfo {volume} {124}},\ \bibinfo {pages} {218301} (\bibinfo {year} {2020})}\BibitemShut {NoStop}%
\bibitem [{\citenamefont {Jalan}\ \emph {et~al.}(2019)\citenamefont {Jalan}, \citenamefont {Rathore}, \citenamefont {Kachhvah},\ and\ \citenamefont {Yadav}}]{jalan2019inhibition}%
  \BibitemOpen
  \bibfield  {author} {\bibinfo {author} {\bibfnamefont {S.}~\bibnamefont {Jalan}}, \bibinfo {author} {\bibfnamefont {V.}~\bibnamefont {Rathore}}, \bibinfo {author} {\bibfnamefont {A.~D.}\ \bibnamefont {Kachhvah}}, \ and\ \bibinfo {author} {\bibfnamefont {A.}~\bibnamefont {Yadav}},\ }\bibfield  {title} {\enquote {\bibinfo {title} {Inhibition-induced explosive synchronization in multiplex networks},}\ }\href@noop {} {\bibfield  {journal} {\bibinfo  {journal} {Physical Review E}\ }\textbf {\bibinfo {volume} {99}},\ \bibinfo {pages} {062305} (\bibinfo {year} {2019})}\BibitemShut {NoStop}%
\bibitem [{\citenamefont {Zhang}\ \emph {et~al.}(2016)\citenamefont {Zhang}, \citenamefont {Guan}, \citenamefont {Zou}, \citenamefont {Chen},\ and\ \citenamefont {Liu}}]{zhang2016suppressing}%
  \BibitemOpen
  \bibfield  {author} {\bibinfo {author} {\bibfnamefont {X.}~\bibnamefont {Zhang}}, \bibinfo {author} {\bibfnamefont {S.}~\bibnamefont {Guan}}, \bibinfo {author} {\bibfnamefont {Y.}~\bibnamefont {Zou}}, \bibinfo {author} {\bibfnamefont {X.}~\bibnamefont {Chen}}, \ and\ \bibinfo {author} {\bibfnamefont {Z.}~\bibnamefont {Liu}},\ }\bibfield  {title} {\enquote {\bibinfo {title} {Suppressing explosive synchronization by contrarians},}\ }\href@noop {} {\bibfield  {journal} {\bibinfo  {journal} {Europhysics Letters}\ }\textbf {\bibinfo {volume} {113}},\ \bibinfo {pages} {28005} (\bibinfo {year} {2016})}\BibitemShut {NoStop}%
\bibitem [{\citenamefont {Rathore}, \citenamefont {Suman},\ and\ \citenamefont {Jalan}(2023)}]{rathore2023synchronization}%
  \BibitemOpen
  \bibfield  {author} {\bibinfo {author} {\bibfnamefont {V.}~\bibnamefont {Rathore}}, \bibinfo {author} {\bibfnamefont {A.}~\bibnamefont {Suman}}, \ and\ \bibinfo {author} {\bibfnamefont {S.}~\bibnamefont {Jalan}},\ }\bibfield  {title} {\enquote {\bibinfo {title} {Synchronization onset for contrarians with higher-order interactions in multilayer systems},}\ }\href@noop {} {\bibfield  {journal} {\bibinfo  {journal} {Chaos: An Interdisciplinary Journal of Nonlinear Science}\ }\textbf {\bibinfo {volume} {33}} (\bibinfo {year} {2023})}\BibitemShut {NoStop}%
\bibitem [{\citenamefont {Soriano}\ \emph {et~al.}(2008)\citenamefont {Soriano}, \citenamefont {Rodr{\'\i}guez~Mart{\'\i}nez}, \citenamefont {Tlusty},\ and\ \citenamefont {Moses}}]{soriano2008development}%
  \BibitemOpen
  \bibfield  {author} {\bibinfo {author} {\bibfnamefont {J.}~\bibnamefont {Soriano}}, \bibinfo {author} {\bibfnamefont {M.}~\bibnamefont {Rodr{\'\i}guez~Mart{\'\i}nez}}, \bibinfo {author} {\bibfnamefont {T.}~\bibnamefont {Tlusty}}, \ and\ \bibinfo {author} {\bibfnamefont {E.}~\bibnamefont {Moses}},\ }\bibfield  {title} {\enquote {\bibinfo {title} {Development of input connections in neural cultures},}\ }\href@noop {} {\bibfield  {journal} {\bibinfo  {journal} {Proceedings of the National Academy of Sciences}\ }\textbf {\bibinfo {volume} {105}},\ \bibinfo {pages} {13758--13763} (\bibinfo {year} {2008})}\BibitemShut {NoStop}%
\bibitem [{\citenamefont {Myung}\ \emph {et~al.}(2015)\citenamefont {Myung}, \citenamefont {Hong}, \citenamefont {DeWoskin}, \citenamefont {De~Schutter}, \citenamefont {Forger},\ and\ \citenamefont {Takumi}}]{myung2015gaba}%
  \BibitemOpen
  \bibfield  {author} {\bibinfo {author} {\bibfnamefont {J.}~\bibnamefont {Myung}}, \bibinfo {author} {\bibfnamefont {S.}~\bibnamefont {Hong}}, \bibinfo {author} {\bibfnamefont {D.}~\bibnamefont {DeWoskin}}, \bibinfo {author} {\bibfnamefont {E.}~\bibnamefont {De~Schutter}}, \bibinfo {author} {\bibfnamefont {D.~B.}\ \bibnamefont {Forger}}, \ and\ \bibinfo {author} {\bibfnamefont {T.}~\bibnamefont {Takumi}},\ }\bibfield  {title} {\enquote {\bibinfo {title} {Gaba-mediated repulsive coupling between circadian clock neurons in the \uppercase{SCN} encodes seasonal time},}\ }\href@noop {} {\bibfield  {journal} {\bibinfo  {journal} {Proceedings of the National Academy of Sciences}\ }\textbf {\bibinfo {volume} {112}},\ \bibinfo {pages} {E3920--E3929} (\bibinfo {year} {2015})}\BibitemShut {NoStop}%
\bibitem [{\citenamefont {Yi}\ \emph {et~al.}(2013)\citenamefont {Yi}, \citenamefont {Baek}, \citenamefont {Zhu},\ and\ \citenamefont {Kim}}]{Yi2013}%
  \BibitemOpen
  \bibfield  {author} {\bibinfo {author} {\bibfnamefont {S.~D.}\ \bibnamefont {Yi}}, \bibinfo {author} {\bibfnamefont {S.~K.}\ \bibnamefont {Baek}}, \bibinfo {author} {\bibfnamefont {C.-P.}\ \bibnamefont {Zhu}}, \ and\ \bibinfo {author} {\bibfnamefont {B.~J.}\ \bibnamefont {Kim}},\ }\bibfield  {title} {\enquote {\bibinfo {title} {Phase transition in a coevolving network of conformist and contrarian voters},}\ }\href {\doibase 10.1103/PhysRevE.87.012806} {\bibfield  {journal} {\bibinfo  {journal} {Phys. Rev. E}\ }\textbf {\bibinfo {volume} {87}},\ \bibinfo {pages} {012806} (\bibinfo {year} {2013})}\BibitemShut {NoStop}%
\bibitem [{\citenamefont {Gir{\'o}n}\ \emph {et~al.}(2016)\citenamefont {Gir{\'o}n}, \citenamefont {Saiz}, \citenamefont {Bacelar}, \citenamefont {Andrade},\ and\ \citenamefont {G{\'o}mez-Gardenes}}]{giron2016synchronization}%
  \BibitemOpen
  \bibfield  {author} {\bibinfo {author} {\bibfnamefont {A.}~\bibnamefont {Gir{\'o}n}}, \bibinfo {author} {\bibfnamefont {H.}~\bibnamefont {Saiz}}, \bibinfo {author} {\bibfnamefont {F.~S.}\ \bibnamefont {Bacelar}}, \bibinfo {author} {\bibfnamefont {R.~F.}\ \bibnamefont {Andrade}}, \ and\ \bibinfo {author} {\bibfnamefont {J.}~\bibnamefont {G{\'o}mez-Gardenes}},\ }\bibfield  {title} {\enquote {\bibinfo {title} {Synchronization unveils the organization of ecological networks with positive and negative interactions},}\ }\href@noop {} {\bibfield  {journal} {\bibinfo  {journal} {Chaos: An Interdisciplinary Journal of Nonlinear Science}\ }\textbf {\bibinfo {volume} {26}} (\bibinfo {year} {2016})}\BibitemShut {NoStop}%
\bibitem [{\citenamefont {Hong}\ and\ \citenamefont {Strogatz}(2011)}]{hong2011kuramoto}%
  \BibitemOpen
  \bibfield  {author} {\bibinfo {author} {\bibfnamefont {H.}~\bibnamefont {Hong}}\ and\ \bibinfo {author} {\bibfnamefont {S.~H.}\ \bibnamefont {Strogatz}},\ }\bibfield  {title} {\enquote {\bibinfo {title} {Kuramoto model of coupled oscillators with positive and negative coupling parameters: An example of conformist and contrarian oscillators},}\ }\href@noop {} {\bibfield  {journal} {\bibinfo  {journal} {Physical Review Letters}\ }\textbf {\bibinfo {volume} {106}},\ \bibinfo {pages} {054102} (\bibinfo {year} {2011})}\BibitemShut {NoStop}%
\bibitem [{\citenamefont {Kumar}\ \emph {et~al.}(2015)\citenamefont {Kumar}, \citenamefont {Verma}, \citenamefont {Parmananda},\ and\ \citenamefont {Boccaletti}}]{kumar2015experimental}%
  \BibitemOpen
  \bibfield  {author} {\bibinfo {author} {\bibfnamefont {P.}~\bibnamefont {Kumar}}, \bibinfo {author} {\bibfnamefont {D.~K.}\ \bibnamefont {Verma}}, \bibinfo {author} {\bibfnamefont {P.}~\bibnamefont {Parmananda}}, \ and\ \bibinfo {author} {\bibfnamefont {S.}~\bibnamefont {Boccaletti}},\ }\bibfield  {title} {\enquote {\bibinfo {title} {Experimental evidence of explosive synchronization in mercury beating-heart oscillators},}\ }\href@noop {} {\bibfield  {journal} {\bibinfo  {journal} {Physical Review E}\ }\textbf {\bibinfo {volume} {91}},\ \bibinfo {pages} {062909} (\bibinfo {year} {2015})}\BibitemShut {NoStop}%
\bibitem [{\citenamefont {Mahler}, \citenamefont {Friesem},\ and\ \citenamefont {Davidson}(2020)}]{mahler2020experimental}%
  \BibitemOpen
  \bibfield  {author} {\bibinfo {author} {\bibfnamefont {S.}~\bibnamefont {Mahler}}, \bibinfo {author} {\bibfnamefont {A.~A.}\ \bibnamefont {Friesem}}, \ and\ \bibinfo {author} {\bibfnamefont {N.}~\bibnamefont {Davidson}},\ }\bibfield  {title} {\enquote {\bibinfo {title} {Experimental demonstration of crowd synchrony and first-order transition with lasers},}\ }\href@noop {} {\bibfield  {journal} {\bibinfo  {journal} {Physical Review Research}\ }\textbf {\bibinfo {volume} {2}},\ \bibinfo {pages} {043220} (\bibinfo {year} {2020})}\BibitemShut {NoStop}%
\bibitem [{\citenamefont {Taylor}\ \emph {et~al.}(2009)\citenamefont {Taylor}, \citenamefont {Tinsley}, \citenamefont {Wang}, \citenamefont {Huang},\ and\ \citenamefont {Showalter}}]{taylor2009dynamical}%
  \BibitemOpen
  \bibfield  {author} {\bibinfo {author} {\bibfnamefont {A.~F.}\ \bibnamefont {Taylor}}, \bibinfo {author} {\bibfnamefont {M.~R.}\ \bibnamefont {Tinsley}}, \bibinfo {author} {\bibfnamefont {F.}~\bibnamefont {Wang}}, \bibinfo {author} {\bibfnamefont {Z.}~\bibnamefont {Huang}}, \ and\ \bibinfo {author} {\bibfnamefont {K.}~\bibnamefont {Showalter}},\ }\bibfield  {title} {\enquote {\bibinfo {title} {Dynamical quorum sensing and synchronization in large populations of chemical oscillators},}\ }\href@noop {} {\bibfield  {journal} {\bibinfo  {journal} {Science}\ }\textbf {\bibinfo {volume} {323}},\ \bibinfo {pages} {614--617} (\bibinfo {year} {2009})}\BibitemShut {NoStop}%
\bibitem [{\citenamefont {Leyva}\ \emph {et~al.}(2012)\citenamefont {Leyva}, \citenamefont {Sevilla-Escoboza}, \citenamefont {Buld{\'u}}, \citenamefont {Sendi{\~n}a-Nadal}, \citenamefont {G{\'o}mez-Gardenes}, \citenamefont {Arenas}, \citenamefont {Moreno}, \citenamefont {G{\'o}mez}, \citenamefont {Jaimes-Reategui},\ and\ \citenamefont {Boccaletti}}]{leyva2012explosive}%
  \BibitemOpen
  \bibfield  {author} {\bibinfo {author} {\bibfnamefont {I.}~\bibnamefont {Leyva}}, \bibinfo {author} {\bibfnamefont {R.}~\bibnamefont {Sevilla-Escoboza}}, \bibinfo {author} {\bibfnamefont {J.}~\bibnamefont {Buld{\'u}}}, \bibinfo {author} {\bibfnamefont {I.}~\bibnamefont {Sendi{\~n}a-Nadal}}, \bibinfo {author} {\bibfnamefont {J.}~\bibnamefont {G{\'o}mez-Gardenes}}, \bibinfo {author} {\bibfnamefont {A.}~\bibnamefont {Arenas}}, \bibinfo {author} {\bibfnamefont {Y.}~\bibnamefont {Moreno}}, \bibinfo {author} {\bibfnamefont {S.}~\bibnamefont {G{\'o}mez}}, \bibinfo {author} {\bibfnamefont {R.}~\bibnamefont {Jaimes-Reategui}}, \ and\ \bibinfo {author} {\bibfnamefont {S.}~\bibnamefont {Boccaletti}},\ }\bibfield  {title} {\enquote {\bibinfo {title} {Explosive first-order transition to synchrony in networked chaotic oscillators},}\ }\href@noop {} {\bibfield  {journal} {\bibinfo  {journal} {Physical Review Letters}\ }\textbf {\bibinfo {volume} {108}},\ \bibinfo {pages} {168702} (\bibinfo {year} {2012})}\BibitemShut
  {NoStop}%
\bibitem [{\citenamefont {C{\u{a}}lug{\u{a}}ru}\ \emph {et~al.}(2020)\citenamefont {C{\u{a}}lug{\u{a}}ru}, \citenamefont {Totz}, \citenamefont {Martens},\ and\ \citenamefont {Engel}}]{cualuguaru2020first}%
  \BibitemOpen
  \bibfield  {author} {\bibinfo {author} {\bibfnamefont {D.}~\bibnamefont {C{\u{a}}lug{\u{a}}ru}}, \bibinfo {author} {\bibfnamefont {J.~F.}\ \bibnamefont {Totz}}, \bibinfo {author} {\bibfnamefont {E.~A.}\ \bibnamefont {Martens}}, \ and\ \bibinfo {author} {\bibfnamefont {H.}~\bibnamefont {Engel}},\ }\bibfield  {title} {\enquote {\bibinfo {title} {First-order synchronization transition in a large population of strongly coupled relaxation oscillators},}\ }\href@noop {} {\bibfield  {journal} {\bibinfo  {journal} {Science Advances}\ }\textbf {\bibinfo {volume} {6}},\ \bibinfo {pages} {eabb2637} (\bibinfo {year} {2020})}\BibitemShut {NoStop}%
\bibitem [{\citenamefont {Pope}(2001)}]{pope2001turbulent}%
  \BibitemOpen
  \bibfield  {author} {\bibinfo {author} {\bibfnamefont {S.~B.}\ \bibnamefont {Pope}},\ }\bibfield  {title} {\enquote {\bibinfo {title} {Turbulent flows},}\ }\href@noop {} {\bibfield  {journal} {\bibinfo  {journal} {Measurement Science and Technology}\ }\textbf {\bibinfo {volume} {12}},\ \bibinfo {pages} {2020--2021} (\bibinfo {year} {2001})}\BibitemShut {NoStop}%
\bibitem [{\citenamefont {Nair}\ and\ \citenamefont {Sujith}(2014)}]{nair2014multifractality}%
  \BibitemOpen
  \bibfield  {author} {\bibinfo {author} {\bibfnamefont {V.}~\bibnamefont {Nair}}\ and\ \bibinfo {author} {\bibfnamefont {R.~I.}\ \bibnamefont {Sujith}},\ }\bibfield  {title} {\enquote {\bibinfo {title} {Multifractality in combustion noise: predicting an impending combustion instability},}\ }\href@noop {} {\bibfield  {journal} {\bibinfo  {journal} {Journal of Fluid Mechanics}\ }\textbf {\bibinfo {volume} {747}},\ \bibinfo {pages} {635--655} (\bibinfo {year} {2014})}\BibitemShut {NoStop}%
\bibitem [{\citenamefont {Sujith}\ and\ \citenamefont {Unni}(2020)}]{sujith2020complex}%
  \BibitemOpen
  \bibfield  {author} {\bibinfo {author} {\bibfnamefont {R.~I.}\ \bibnamefont {Sujith}}\ and\ \bibinfo {author} {\bibfnamefont {V.~R.}\ \bibnamefont {Unni}},\ }\bibfield  {title} {\enquote {\bibinfo {title} {Complex system approach to investigate and mitigate thermoacoustic instability in turbulent combustors},}\ }\href@noop {} {\bibfield  {journal} {\bibinfo  {journal} {Physics of Fluids}\ }\textbf {\bibinfo {volume} {32}},\ \bibinfo {pages} {061401} (\bibinfo {year} {2020})}\BibitemShut {NoStop}%
\bibitem [{\citenamefont {Juniper}\ and\ \citenamefont {Sujith}(2018)}]{juniper2018sensitivity}%
  \BibitemOpen
  \bibfield  {author} {\bibinfo {author} {\bibfnamefont {M.~P.}\ \bibnamefont {Juniper}}\ and\ \bibinfo {author} {\bibfnamefont {R.~I.}\ \bibnamefont {Sujith}},\ }\bibfield  {title} {\enquote {\bibinfo {title} {Sensitivity and nonlinearity of thermoacoustic oscillations},}\ }\href@noop {} {\bibfield  {journal} {\bibinfo  {journal} {Annual Review of Fluid Mechanics}\ }\textbf {\bibinfo {volume} {50}},\ \bibinfo {pages} {661--689} (\bibinfo {year} {2018})}\BibitemShut {NoStop}%
\bibitem [{\citenamefont {Biggs}(2009)}]{biggs2009rocketdyne}%
  \BibitemOpen
  \bibfield  {author} {\bibinfo {author} {\bibfnamefont {R.}~\bibnamefont {Biggs}},\ }\href@noop {} {\emph {\bibinfo {title} {``Rocketdyne-F-1 Saturn V first stage engine'' in Remembering the Giants: Apollo Rocket Propulsion Development}}},\ edited by\ \bibinfo {editor} {\bibfnamefont {S.~C.}\ \bibnamefont {Fisher}}\ and\ \bibinfo {editor} {\bibfnamefont {S.~A.}\ \bibnamefont {Rahman}}\ (\bibinfo  {publisher} {NASA},\ \bibinfo {year} {2009})\ pp.\ \bibinfo {pages} {15--26}\BibitemShut {NoStop}%
\bibitem [{\citenamefont {Lieuwen}\ and\ \citenamefont {Yang}(2005)}]{lieuwen2005combustion}%
  \BibitemOpen
  \bibfield  {author} {\bibinfo {author} {\bibfnamefont {T.~C.}\ \bibnamefont {Lieuwen}}\ and\ \bibinfo {author} {\bibfnamefont {V.}~\bibnamefont {Yang}},\ }\href@noop {} {\emph {\bibinfo {title} {Combustion instabilities in gas turbine engines: operational experience, fundamental mechanisms, and modeling}}}\ (\bibinfo  {publisher} {American Institute of Aeronautics and Astronautics},\ \bibinfo {year} {2005})\BibitemShut {NoStop}%
\bibitem [{\citenamefont {Kriesels}\ \emph {et~al.}(1995)\citenamefont {Kriesels}, \citenamefont {Peters}, \citenamefont {Hirschberg}, \citenamefont {Wijnands}, \citenamefont {Iafrati}, \citenamefont {Riccardi}, \citenamefont {Piva},\ and\ \citenamefont {Bruggeman}}]{kriesels1995high}%
  \BibitemOpen
  \bibfield  {author} {\bibinfo {author} {\bibfnamefont {P.}~\bibnamefont {Kriesels}}, \bibinfo {author} {\bibfnamefont {M.}~\bibnamefont {Peters}}, \bibinfo {author} {\bibfnamefont {A.}~\bibnamefont {Hirschberg}}, \bibinfo {author} {\bibfnamefont {A.}~\bibnamefont {Wijnands}}, \bibinfo {author} {\bibfnamefont {A.}~\bibnamefont {Iafrati}}, \bibinfo {author} {\bibfnamefont {G.}~\bibnamefont {Riccardi}}, \bibinfo {author} {\bibfnamefont {R.}~\bibnamefont {Piva}}, \ and\ \bibinfo {author} {\bibfnamefont {J.}~\bibnamefont {Bruggeman}},\ }\bibfield  {title} {\enquote {\bibinfo {title} {High amplitude vortex-induced pulsations in a gas transport system},}\ }\href@noop {} {\bibfield  {journal} {\bibinfo  {journal} {Journal of Sound and Vibration}\ }\textbf {\bibinfo {volume} {184}},\ \bibinfo {pages} {343--368} (\bibinfo {year} {1995})}\BibitemShut {NoStop}%
\bibitem [{\citenamefont {Green}\ and\ \citenamefont {Unruh}(2006)}]{green2006failure}%
  \BibitemOpen
  \bibfield  {author} {\bibinfo {author} {\bibfnamefont {D.}~\bibnamefont {Green}}\ and\ \bibinfo {author} {\bibfnamefont {W.~G.}\ \bibnamefont {Unruh}},\ }\bibfield  {title} {\enquote {\bibinfo {title} {The failure of the \uppercase{T}acoma bridge: A physical model},}\ }\href@noop {} {\bibfield  {journal} {\bibinfo  {journal} {American Journal of Physics}\ }\textbf {\bibinfo {volume} {74}},\ \bibinfo {pages} {706--716} (\bibinfo {year} {2006})}\BibitemShut {NoStop}%
\bibitem [{\citenamefont {Pavithran}\ \emph {et~al.}(2020{\natexlab{a}})\citenamefont {Pavithran}, \citenamefont {Unni}, \citenamefont {Varghese}, \citenamefont {Sujith}, \citenamefont {Saha}, \citenamefont {Marwan},\ and\ \citenamefont {Kurths}}]{pavithran2020universality_inst}%
  \BibitemOpen
  \bibfield  {author} {\bibinfo {author} {\bibfnamefont {I.}~\bibnamefont {Pavithran}}, \bibinfo {author} {\bibfnamefont {V.~R.}\ \bibnamefont {Unni}}, \bibinfo {author} {\bibfnamefont {A.~J.}\ \bibnamefont {Varghese}}, \bibinfo {author} {\bibfnamefont {R.~I.}\ \bibnamefont {Sujith}}, \bibinfo {author} {\bibfnamefont {A.}~\bibnamefont {Saha}}, \bibinfo {author} {\bibfnamefont {N.}~\bibnamefont {Marwan}}, \ and\ \bibinfo {author} {\bibfnamefont {J.}~\bibnamefont {Kurths}},\ }\bibfield  {title} {\enquote {\bibinfo {title} {Universality in the emergence of oscillatory instabilities in turbulent flows},}\ }\href@noop {} {\bibfield  {journal} {\bibinfo  {journal} {Europhysics Letters}\ }\textbf {\bibinfo {volume} {129}},\ \bibinfo {pages} {24004} (\bibinfo {year} {2020}{\natexlab{a}})}\BibitemShut {NoStop}%
\bibitem [{\citenamefont {Pavithran}\ \emph {et~al.}(2020{\natexlab{b}})\citenamefont {Pavithran}, \citenamefont {Unni}, \citenamefont {Varghese}, \citenamefont {Premraj}, \citenamefont {Sujith}, \citenamefont {Vijayan}, \citenamefont {Saha}, \citenamefont {Marwan},\ and\ \citenamefont {Kurths}}]{pavithran2020universality_spectral}%
  \BibitemOpen
  \bibfield  {author} {\bibinfo {author} {\bibfnamefont {I.}~\bibnamefont {Pavithran}}, \bibinfo {author} {\bibfnamefont {V.~R.}\ \bibnamefont {Unni}}, \bibinfo {author} {\bibfnamefont {A.~J.}\ \bibnamefont {Varghese}}, \bibinfo {author} {\bibfnamefont {D.}~\bibnamefont {Premraj}}, \bibinfo {author} {\bibfnamefont {R.~I.}\ \bibnamefont {Sujith}}, \bibinfo {author} {\bibfnamefont {C.}~\bibnamefont {Vijayan}}, \bibinfo {author} {\bibfnamefont {A.}~\bibnamefont {Saha}}, \bibinfo {author} {\bibfnamefont {N.}~\bibnamefont {Marwan}}, \ and\ \bibinfo {author} {\bibfnamefont {J.}~\bibnamefont {Kurths}},\ }\bibfield  {title} {\enquote {\bibinfo {title} {Universality in spectral condensation},}\ }\href@noop {} {\bibfield  {journal} {\bibinfo  {journal} {Scientific Reports}\ }\textbf {\bibinfo {volume} {10}},\ \bibinfo {pages} {17405} (\bibinfo {year} {2020}{\natexlab{b}})}\BibitemShut {NoStop}%
\bibitem [{\citenamefont {Krishnan}\ \emph {et~al.}(2019)\citenamefont {Krishnan}, \citenamefont {Sujith}, \citenamefont {Marwan},\ and\ \citenamefont {Kurths}}]{krishnan2019emergence}%
  \BibitemOpen
  \bibfield  {author} {\bibinfo {author} {\bibfnamefont {A.}~\bibnamefont {Krishnan}}, \bibinfo {author} {\bibfnamefont {R.~I.}\ \bibnamefont {Sujith}}, \bibinfo {author} {\bibfnamefont {N.}~\bibnamefont {Marwan}}, \ and\ \bibinfo {author} {\bibfnamefont {J.}~\bibnamefont {Kurths}},\ }\bibfield  {title} {\enquote {\bibinfo {title} {On the emergence of large clusters of acoustic power sources at the onset of thermoacoustic instability in a turbulent combustor},}\ }\href@noop {} {\bibfield  {journal} {\bibinfo  {journal} {Journal of Fluid Mechanics}\ }\textbf {\bibinfo {volume} {874}},\ \bibinfo {pages} {455--482} (\bibinfo {year} {2019})}\BibitemShut {NoStop}%
\bibitem [{\citenamefont {Joseph}, \citenamefont {Pavithran},\ and\ \citenamefont {Sujith}(2024)}]{joseph2024explosive}%
  \BibitemOpen
  \bibfield  {author} {\bibinfo {author} {\bibfnamefont {A.}~\bibnamefont {Joseph}}, \bibinfo {author} {\bibfnamefont {I.}~\bibnamefont {Pavithran}}, \ and\ \bibinfo {author} {\bibfnamefont {R.~I.}\ \bibnamefont {Sujith}},\ }\bibfield  {title} {\enquote {\bibinfo {title} {Explosive synchronization in a turbulent reactive flow system},}\ }\href@noop {} {\bibfield  {journal} {\bibinfo  {journal} {Chaos: An Interdisciplinary Journal of Nonlinear Science}\ }\textbf {\bibinfo {volume} {34}} (\bibinfo {year} {2024})}\BibitemShut {NoStop}%
\bibitem [{\citenamefont {Rayleigh}(1878)}]{rayleigh1878explanation}%
  \BibitemOpen
  \bibfield  {author} {\bibinfo {author} {\bibfnamefont {L.}~\bibnamefont {Rayleigh}},\ }\bibfield  {title} {\enquote {\bibinfo {title} {The explanation of certain acoustical phenomena},}\ }\href@noop {} {\bibfield  {journal} {\bibinfo  {journal} {Nature}\ }\textbf {\bibinfo {volume} {18}},\ \bibinfo {pages} {319–321} (\bibinfo {year} {1878})}\BibitemShut {NoStop}%
\bibitem [{\citenamefont {Poinsot}\ and\ \citenamefont {Veynante}(2005)}]{poinsot2005theoretical}%
  \BibitemOpen
  \bibfield  {author} {\bibinfo {author} {\bibfnamefont {T.}~\bibnamefont {Poinsot}}\ and\ \bibinfo {author} {\bibfnamefont {D.}~\bibnamefont {Veynante}},\ }\href@noop {} {\emph {\bibinfo {title} {Theoretical and numerical combustion}}}\ (\bibinfo  {publisher} {RT Edwards, Inc.},\ \bibinfo {year} {2005})\BibitemShut {NoStop}%
\bibitem [{\citenamefont {Samaniego}\ \emph {et~al.}(1993)\citenamefont {Samaniego}, \citenamefont {Yip}, \citenamefont {Poinsot},\ and\ \citenamefont {Candel}}]{samaniego1993low}%
  \BibitemOpen
  \bibfield  {author} {\bibinfo {author} {\bibfnamefont {J.}~\bibnamefont {Samaniego}}, \bibinfo {author} {\bibfnamefont {B.}~\bibnamefont {Yip}}, \bibinfo {author} {\bibfnamefont {T.}~\bibnamefont {Poinsot}}, \ and\ \bibinfo {author} {\bibfnamefont {S.}~\bibnamefont {Candel}},\ }\bibfield  {title} {\enquote {\bibinfo {title} {Low-frequency combustion instability mechanisms in a side-dump combustor},}\ }\href@noop {} {\bibfield  {journal} {\bibinfo  {journal} {Combustion and Flame}\ }\textbf {\bibinfo {volume} {94}},\ \bibinfo {pages} {363--380} (\bibinfo {year} {1993})}\BibitemShut {NoStop}%
\bibitem [{\citenamefont {Tony}\ \emph {et~al.}(2015)\citenamefont {Tony}, \citenamefont {Gopalakrishnan}, \citenamefont {Sreelekha},\ and\ \citenamefont {Sujith}}]{tony2015detecting}%
  \BibitemOpen
  \bibfield  {author} {\bibinfo {author} {\bibfnamefont {J.}~\bibnamefont {Tony}}, \bibinfo {author} {\bibfnamefont {E.~A.}\ \bibnamefont {Gopalakrishnan}}, \bibinfo {author} {\bibfnamefont {E.}~\bibnamefont {Sreelekha}}, \ and\ \bibinfo {author} {\bibfnamefont {R.~I.}\ \bibnamefont {Sujith}},\ }\bibfield  {title} {\enquote {\bibinfo {title} {Detecting deterministic nature of pressure measurements from a turbulent combustor},}\ }\href@noop {} {\bibfield  {journal} {\bibinfo  {journal} {Physical Review E}\ }\textbf {\bibinfo {volume} {92}},\ \bibinfo {pages} {062902} (\bibinfo {year} {2015})}\BibitemShut {NoStop}%
\bibitem [{\citenamefont {Pawar}\ \emph {et~al.}(2017)\citenamefont {Pawar}, \citenamefont {Seshadri}, \citenamefont {Unni},\ and\ \citenamefont {Sujith}}]{pawar2017thermoacoustic}%
  \BibitemOpen
  \bibfield  {author} {\bibinfo {author} {\bibfnamefont {S.~A.}\ \bibnamefont {Pawar}}, \bibinfo {author} {\bibfnamefont {A.}~\bibnamefont {Seshadri}}, \bibinfo {author} {\bibfnamefont {V.~R.}\ \bibnamefont {Unni}}, \ and\ \bibinfo {author} {\bibfnamefont {R.~I.}\ \bibnamefont {Sujith}},\ }\bibfield  {title} {\enquote {\bibinfo {title} {Thermoacoustic instability as mutual synchronization between the acoustic field of the confinement and turbulent reactive flow},}\ }\href@noop {} {\bibfield  {journal} {\bibinfo  {journal} {Journal of Fluid Mechanics}\ }\textbf {\bibinfo {volume} {827}},\ \bibinfo {pages} {664--693} (\bibinfo {year} {2017})}\BibitemShut {NoStop}%
\bibitem [{\citenamefont {Haken}(1977)}]{haken1977synergetics}%
  \BibitemOpen
  \bibfield  {author} {\bibinfo {author} {\bibfnamefont {H.}~\bibnamefont {Haken}},\ }\bibfield  {title} {\enquote {\bibinfo {title} {Synergetics},}\ }\href@noop {} {\bibfield  {journal} {\bibinfo  {journal} {Physics Bulletin}\ }\textbf {\bibinfo {volume} {28}},\ \bibinfo {pages} {412} (\bibinfo {year} {1977})}\BibitemShut {NoStop}%
\bibitem [{\citenamefont {Bird}\ \emph {et~al.}(1961)\citenamefont {Bird}, \citenamefont {Stewart}, \citenamefont {Lightfoot},\ and\ \citenamefont {Meredith}}]{bird1961transport}%
  \BibitemOpen
  \bibfield  {author} {\bibinfo {author} {\bibfnamefont {R.~B.}\ \bibnamefont {Bird}}, \bibinfo {author} {\bibfnamefont {W.~E.}\ \bibnamefont {Stewart}}, \bibinfo {author} {\bibfnamefont {E.~N.}\ \bibnamefont {Lightfoot}}, \ and\ \bibinfo {author} {\bibfnamefont {R.~E.}\ \bibnamefont {Meredith}},\ }\bibfield  {title} {\enquote {\bibinfo {title} {Transport phenomena},}\ }\href@noop {} {\bibfield  {journal} {\bibinfo  {journal} {Journal of The Electrochemical Society}\ }\textbf {\bibinfo {volume} {108}},\ \bibinfo {pages} {78C} (\bibinfo {year} {1961})}\BibitemShut {NoStop}%
\bibitem [{\citenamefont {Hardalupas}\ and\ \citenamefont {Orain}(2004)}]{hardalupas2004local}%
  \BibitemOpen
  \bibfield  {author} {\bibinfo {author} {\bibfnamefont {Y.}~\bibnamefont {Hardalupas}}\ and\ \bibinfo {author} {\bibfnamefont {M.}~\bibnamefont {Orain}},\ }\bibfield  {title} {\enquote {\bibinfo {title} {Local measurements of the time-dependent heat release rate and equivalence ratio using chemiluminescent emission from a flame},}\ }\href@noop {} {\bibfield  {journal} {\bibinfo  {journal} {Combustion and Flame}\ }\textbf {\bibinfo {volume} {139}},\ \bibinfo {pages} {188--207} (\bibinfo {year} {2004})}\BibitemShut {NoStop}%
\bibitem [{\citenamefont {Guethe}\ \emph {et~al.}(2012)\citenamefont {Guethe}, \citenamefont {Guyot}, \citenamefont {Singla}, \citenamefont {Noiray},\ and\ \citenamefont {Schuermans}}]{guethe2012chemiluminescence}%
  \BibitemOpen
  \bibfield  {author} {\bibinfo {author} {\bibfnamefont {F.}~\bibnamefont {Guethe}}, \bibinfo {author} {\bibfnamefont {D.}~\bibnamefont {Guyot}}, \bibinfo {author} {\bibfnamefont {G.}~\bibnamefont {Singla}}, \bibinfo {author} {\bibfnamefont {N.}~\bibnamefont {Noiray}}, \ and\ \bibinfo {author} {\bibfnamefont {B.}~\bibnamefont {Schuermans}},\ }\bibfield  {title} {\enquote {\bibinfo {title} {Chemiluminescence as diagnostic tool in the development of gas turbines},}\ }\href@noop {} {\bibfield  {journal} {\bibinfo  {journal} {Applied Physics B}\ }\textbf {\bibinfo {volume} {107}},\ \bibinfo {pages} {619--636} (\bibinfo {year} {2012})}\BibitemShut {NoStop}%
\bibitem [{\citenamefont {Schadow}\ and\ \citenamefont {Gutmark}(1992)}]{schadow1992combustion}%
  \BibitemOpen
  \bibfield  {author} {\bibinfo {author} {\bibfnamefont {K.~C.}\ \bibnamefont {Schadow}}\ and\ \bibinfo {author} {\bibfnamefont {E.}~\bibnamefont {Gutmark}},\ }\bibfield  {title} {\enquote {\bibinfo {title} {Combustion instability related to vortex shedding in dump combustors and their passive control},}\ }\href@noop {} {\bibfield  {journal} {\bibinfo  {journal} {Progress in Energy and Combustion Science}\ }\textbf {\bibinfo {volume} {18}},\ \bibinfo {pages} {117--132} (\bibinfo {year} {1992})}\BibitemShut {NoStop}%
\end{thebibliography}%

\section{Supporting Information}

% \subsection*{Acoustic pressure variation across the region of interest}
% The acoustic pressure fluctuations measured at three locations ($x=120$, $230$, and $460$ mm) along the longitudinal axis of the turbulent reactive flow system are shown in Fig.~\ref{Fig: pressure spatial}.  The pressure fluctuations recorded at these three locations are almost identical. The region of interest for our study is a square section of 30 $\times$ 30 mm as indicated in Fig.~\ref{fig: experimentalSetup}. Further, the distance between the pressure measurement locations is greater than the size of the region of interest (30 mm). Therefore, the acoustic pressure fluctuations is uniform over the region of interest. 

% \begin{figure}[h]
% \centering
% \includegraphics[width=0.5\linewidth]{RESULTS/pressure_2.png}
% \caption{
% \doublespacing
% \footnotesize
% The acoustic pressure fluctuations normalized with their standard deviations for three different spatial locations along the longitudinal axis of the combustor at $x$= 120 mm ($p_1$), 230 mm ($p_2$), and 460 mm ($p_3$) for the chaotic state at $Re=2.45 \times 10^4$ (a) and for periodic oscillatory state (b) at $Re=3.72 \times 10^4$. (c) The rms of the acoustic pressure for $Re=3.72 \times 10^4$ along the axial direction.  
% }
% \label{Fig: pressure spatial}
% \end{figure}

\subsection{Dependence of the window size of the time series}
\begin{figure}
    \centering
    \includegraphics[width=\linewidth]{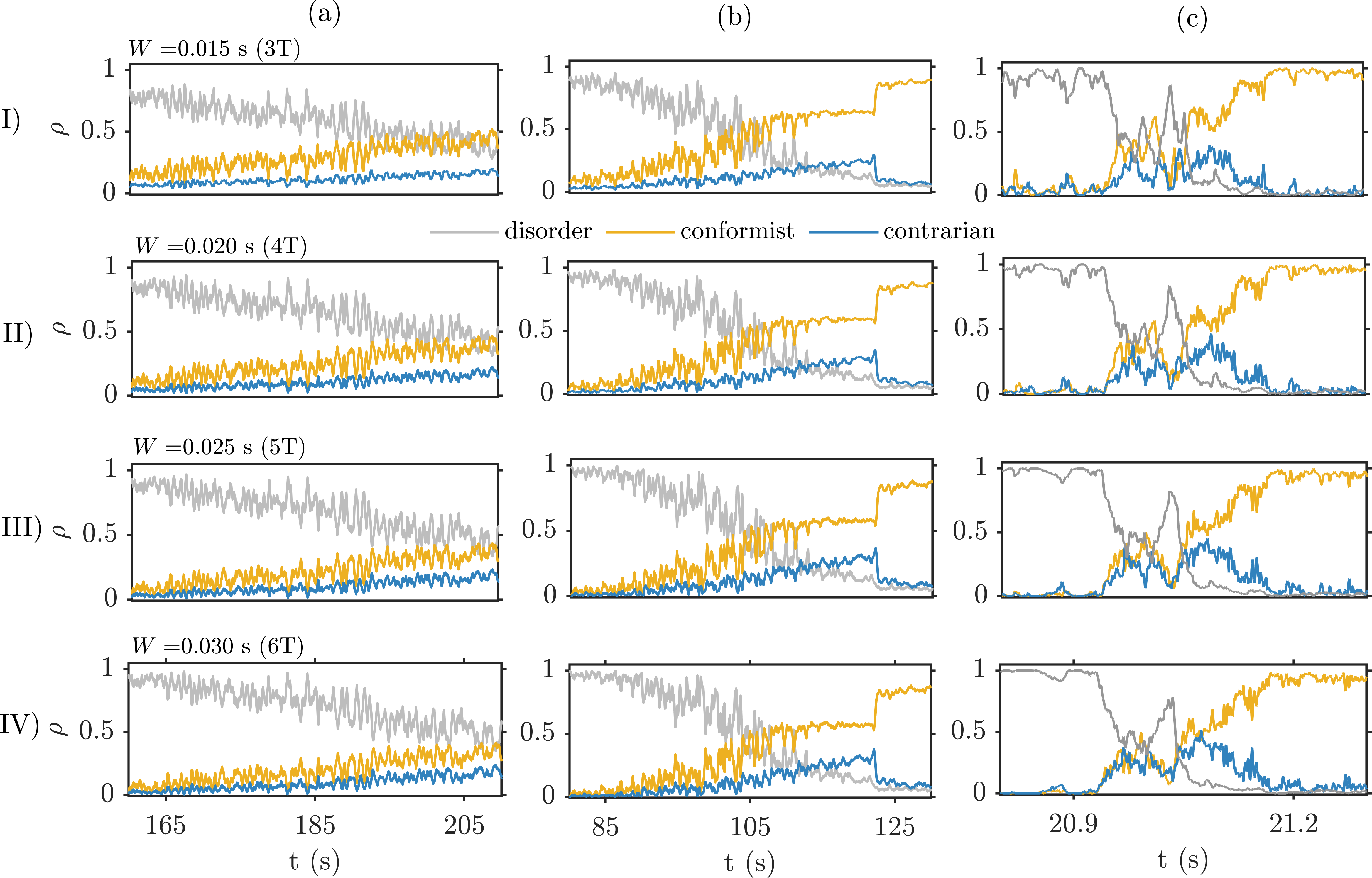}   
    \caption{
    {a-e}, The evolution of the order parameters for conformist, contrarian, and disordered dynamics for different window sizes of ${W}=3T$, $4T$, $5T$, and $6T$. Our results qualitatively remain the same for different window sizes considered.}
    \label{fig: windowSize}
\end{figure}

In this section, we discuss the effect of choosing different window sizes for classifying the dynamics between the heat release rate and the acoustic pressure oscillations. We have selected window sizes ranging from $W=3T$ to $W=6T$, where $T$ is the time period of the dominant mode of the acoustic pressure oscillations. We have calculated the order parameters for the conformist, contrarian, and disordered dynamics 
for these different window sizes. For all the window sizes we have considered, we observe a significant presence of disordered dynamics for the smooth transition. Further, accompanying the abrupt transition, the disordered dynamics have disappeared from the spatiotemporal domain. Therefore, for the different window sizes, the qualitative variation of the order parameters remains the same during (1) a smooth transition, (2) a smooth transition followed by an abrupt transition, and (3) an abrupt transition (Fig.~\ref{fig: windowSize}).  

\subsection{Dependence of the threshold value}
Here, we discuss the effect of selecting a different threshold value on our results. For this purpose, we have identified the correlated and the uncorrelated dynamics for different threshold values ($R_{\mathrm{th}}$) of 0.55, 0.6, 0.65 and 0.7. The order parameters are then calculated for the conformist, contrarian, and disordered dynamics for different threshold values. We observe that, for all the threshold values we considered, the evolution of the order parameters, $\rho_{\mathrm{conf}}$, $\rho_{\mathrm{cont}}$, and $\rho_{\mathrm{dis}}$ remain qualitatively the same (Fig.~\ref{fig: windowSize}) for smooth and abrupt transitions.

\end{document}